\documentclass[11pt,journal,onecolumn]{IEEEtran}
\usepackage{graphicx,color}
\usepackage{xspace,subfigure,epsfig,cuted}
\usepackage{xspace}
\usepackage{graphicx,color}
\usepackage{times,amsfonts,amssymb,amsmath,setspace}
\usepackage{setspace}
\usepackage{comment,url,enumerate}
\usepackage{lipsum}
\usepackage{multicol}
\usepackage[noend]{algpseudocode}
\usepackage{float}
\usepackage{varwidth}

\DeclareMathAlphabet\mathbfcal{OMS}{cmsy}{b}{n}
\usepackage{bbm}\usepackage{mathrsfs}\usepackage{float}
\usepackage{cases}%\usepackage{pseudocode}
\usepackage[plain]{algorithm}

\usepackage{graphicx,url,bm,subfigure,comment}
\usepackage{epsfig,textcomp}
\usepackage{times,amsfonts,amssymb,amsmath,setspace}%,cleveref}

\newtheorem{theorem}{Theorem}
\newtheorem{lemma}{Lemma}
\newtheorem{proposition}{Proposition}

\newcommand{\pf}{\noindent{\bf Proof:~}}
\newcommand{\qedsymb}{\hfill{\rule{2mm}{2mm}}}
\DeclareMathAlphabet{\mathsfbf}{OT1}{cmss}{sbc}{n}

\newcommand{\example}[2]{
\begin{center}
\parbox{0.9\columnwidth}{
\rule{0.9\columnwidth}{0.5mm}\\
\noindent {\bf Example~#1:} 
#2\\
\rule{0.9\columnwidth}{0.5mm}
}
\end{center}
}

\newcommand{\RR}{\mathbb{R}} % real set
\newcommand{\ee}{{\rm e}}
\newcommand{\dd}{{\rm\,d}} % differential (for integrals)

\newcommand{\av}{{\bf a}}
\newcommand{\bv}{{\bf b}}

\newcommand{\dv}{{\bf d}}
\newcommand{\ev}{{\bf e}}
\newcommand{\fv}{{\bf f}}

\newcommand{\hv}{{\bf h}}

\newcommand{\mv}{{\bf m}}

\newcommand{\uv}{{\bf u}}
\newcommand{\wv}{{\bf w}}

\newcommand{\xv}{{\bf x}}
\newcommand{\yv}{{\bf y}}
\newcommand{\zv}{{\bf z}}

\newcommand{\Am}{{\bf A}}

\newcommand{\Cm}{{\bf C}}
\newcommand{\Dm}{{\bf D}}

\newcommand{\Hm}{{\bf H}}
\newcommand{\Id}{{\bf I}}

\newcommand{\Lm}{{\bf L}}

\newcommand{\Pm}{{\bf P}}
\newcommand{\Qm}{{\bf Q}}

\newcommand{\Tm}{{\bf T}}

\newcommand{\Wm}{{\bf W}}

\newcommand{\Xm}{{\bf X}}

\newcommand{\Zm}{{\bf Z}}

\newcommand{\Ac}{{\cal A}}
\newcommand{\Bc}{{\cal B}}
\newcommand{\Cc}{{\cal C}}

\newcommand{\Gc}{{\cal G}}

\newcommand{\Ic}{{\cal I}}

\newcommand{\Lc}{{\cal L}}

\newcommand{\Pc}{{\cal P}}

\newcommand{\Uc}{{\cal U}}

\newcommand{\Xc}{{\cal X}}
\newcommand{\Yc}{{\cal Y}}

\newcommand{\betav}{\boldsymbol{\beta}}
\newcommand{\gammav}{\boldsymbol{\gamma}}

\newcommand{\nuv}{\boldsymbol{\nu}}
\newcommand{\muv}{\boldsymbol{\mu }}

\newcommand{\phiv}{\boldsymbol{\phi}}

\newcommand{\Gammam}{\boldsymbol{\Gamma}}

\newcommand{\Sigmam}{\boldsymbol{\Sigma}}

\newcommand{\Psim}{\boldsymbol{\Psi}}
\newcommand{\Thetam}{\boldsymbol{\Theta}}

\newcommand{\Xim}{\boldsymbol{\Xi}}

\newcommand{\diag}{{\hbox{diag}}}

\def\Trace{\mathsf{Tr}}

\def\Tran{^\mathsf{T}}

\def\ben{\begin{enumerate}}
\def\beq{\begin{equation}}
\def\beqa{\begin{eqnarray}}
\def\bit{\begin{itemize}}
\def\een{\end{enumerate}}
\def\eeq{\end{equation}}
\def\eeqa{\end{eqnarray}}
\def\eit{\end{itemize}}

\def\non{\nonumber\\}

\newcommand{\plus}{\mathord{+}}
\newcommand{\minus}{\mathord{-}}

\newcommand{\ls}[1]
   {\dimen0=\fontdimen6\the\font
    \lineskip=#1\dimen0
    \advance\lineskip.5\fontdimen5\the\font
    \advance\lineskip-\dimen0
    \lineskiplimit=.9\lineskip
    \baselineskip=\lineskip
    \advance\baselineskip\dimen0
    \normallineskip\lineskip
    \normallineskiplimit\lineskiplimit
    \normalbaselineskip\baselineskip
    \ignorespaces
}
\makeatother

\begin{document}
\title{Opinion Dynamics on Correlated Subjects\\ in Social Networks}

\author{Alessandro Nordio,~\IEEEmembership{Member,~IEEE,}
  Alberto Tarable,~\IEEEmembership{Member,~IEEE,}
  Carla Fabiana Chiasserini,~\IEEEmembership{Fellow,~IEEE,}
Emilio Leonardi,~\IEEEmembership{Senior Member,~IEEE}\\
\thanks{}%}% <-this % stops a space
\IEEEcompsocitemizethanks{\IEEEcompsocthanksitem A.~Nordio and A. Tarable are with IEIIT-CNR (Institute of Electronics,
  Telecommunications and Information Engineering of the National Research Council of Italy),
  Italy, email: {firstname.lastname}@ieiit.cnr.it. 

\IEEEcompsocthanksitem C. F. Chiasserini and E. Leonardi are with DET,
  Politecnico di Torino, Torino, Italy, email: {firstname.lastname}@polito.it, and are associate researchers with IEIIT-CNR.}}%

\markboth{Journal of \LaTeX\ Class Files,~Vol.~14, No.~8, August~2015}%
{Shell \MakeLowercase{\textit{et al.}}: Bare Demo of IEEEtran.cls for Computer Society Journals}

\IEEEtitleabstractindextext{%
\begin{abstract}
Understanding the evolution of collective beliefs is of critical
  importance to get insights on the political trends as well as on
  social tastes and opinions. In particular, it is pivotal to develop
  analytical models that can predict the beliefs dynamics and capture
  the interdependence of opinions on different subjects. In this paper
  we tackle this issue also accounting for the individual
  endogenous process of opinion evolution, as well as repulsive
  interactions between individuals' opinions that may arise in the
  presence of an adversarial attitude of the  
  individuals. Using a mean field approach, we characterize the time
  evolution of opinions of a large population of individuals through
  a multidimensional Fokker-Planck equation,  and we identify the
  conditions under which stability holds.  Finally, we derive the
  steady-state opinion distribution as a function of the individuals' 
  personality and of the existing social interactions. Our numerical
  results show interesting dynamics in the collective beliefs of different
  social communities, and they highlight the effect of correlated subjects
  as well as of individuals with an adversarial attitude.
\end{abstract}

}

\maketitle

\IEEEdisplaynontitleabstractindextext

% \fi
\IEEEpeerreviewmaketitle

\IEEEraisesectionheading{\section{Introduction}\label{sec:introduction}}

%{\color{red} Aggiungere esempio come il seguente:  Consider a group of people discussing two related
%topics, e.g., fish (as a part of diet) in general and salmon. Salmon
%is nested in fish. A person disliking fish also dislikes salmon. If
%the influence process changes individuals attitudes toward fish,
%say promoting fish as a healthy part of a diet, then the door is
%opened for influences on salmon as a part of this diet. If, on the
%other hand, the influence process changes individuals attitudes
%against fish, say warning that fish are now contaminated by toxic
%chemicals, then the door is closed for influences on salmon as
%part of this diet.}

An increasing deal of attention has recently been devoted to the
understanding and the analysis of collective social belief dynamics
over social networks~\cite{Colbaugh2010,Asur:2010,shi2013agreement,Baccelli}.
This interest has been stimulated by the growing awareness of the
fundamental role that social networks and media may play in the
formation and diffusion of opinions/beliefs.  For example, it is
widely recognized that social media have played a pivotal role in
several recent political events, such as the ``Arabian Spring''
or the last US presidential campaign.  Moreover, the availability of a 
large amount of social data generated by users has attracted the
interest of companies and government agencies, which envision
opportunities for exploiting such data to get important real-time
insights on evolution of trends, tastes, and opinions in the society.

Several experimental approaches, based on sentiment analysis \cite{Pang2008}, have
been proposed for a timely analysis of social dynamics. Furthermore, 
several analytic frameworks  have been developed   with the goal of  understanding  and predicting
dominant belief dynamics. These models aim at providing  important
insights on the dynamics of social interactions, as well as possible explanatory
mechanisms for  the emergence of strong collective opinions.  Additionally, they
have also been used to devise possible efficient strategies to influence social
beliefs.

The existing models can be coarsely partitioned into two classes:
\begin{itemize}
\item Discrete models, in which a discrete variable is associated to
  every individual corresponding to a node on a graph, and represents
  the current belief/position of each individual, e.g., favorable,
  contrary, neutral, with respect to the considered subject.  This
  body of work also includes studies such as~\cite{naming1,naming2}
  where the naming game is used to model phenomena such as opinion
  dynamics in a population of agents.  The social interactions are
  represented by the graph edges and the state of a node changes for
  effect of the interactions with its neighbors, i.e., the state of a
  node is a deterministic/stochastic function of the states of its
  neighboring nodes.  Several different mechanisms, such as the Voter
  Model \cite{liggett1997}, Bootstrap-percolation \cite{Bootstrap},
  and Linear Threshold Models \cite{kempe}, have been proposed.  The
  dynamics of the process terminates when the system reaches a
  globally consistent configuration.

\item Continuous models, in which the opinion of individuals on a
  particular subject is described by means of a continuous variable,
  whose value is adapted as a result of social interactions with
  individuals having different opinions \cite{F&J,Tempo,
    Fagnani-Como,Garnier,9-Blondel,16-Weisbuch,17-Bhattacharyya,noi-TNSE}. Also
  in this case, social interactions between individuals are typically
  modeled by using  (static or dynamic) graphs, which reflect the
  structure of the society and describe how individuals
  interact. 
\end{itemize}

All of the above pieces of work have considered that
 the  beliefs of an individual depend on her social interaction and vary for
  effect of pairwise ``attractive'' forces. In particular, 
a sub-class of continuous models that have attained considerable
popularity, considers the so-called bounded confidence, according to
which interactions between individuals are effective only if their
beliefs are sufficiently close to each
other~\cite{7-Deffuant,8-Hegselmann,degroot1974reaching,Fagnani-Acemoglu,Fagnani-Como,Garnier,9-Blondel,16-Weisbuch,17-Bhattacharyya}. 
%\cite{7-Deffuant,8-Hegselmann}. On the one hand, bounded-confidence models are appealing because they capture the
%aversion to interacting with individuals having very different
%beliefs (e.g., social behaviors such as homophily).  On the other hand, the
%introduction of bounded-confidence poses several challenges because
%the equations driving the individuals' interactions become non linear
%\cite{degroot1974reaching,Fagnani-Acemoglu,Fagnani-Como,Garnier,9-Blondel,16-Weisbuch,17-Bhattacharyya}. Therefore,
%an analysis of the properties of such models appears prohibitive in
%most of the cases. 
Furthermore, all previous models represent the evolution of the individuals'
opinions about a specific subject, neglecting how beliefs on different,
yet correlated, topics may vary over time.  This is essentially
equivalent to assume the evolution of opinions on different subjects
to be independent.  Unfortunately, things are much more involved, and
opinions on correlated topics exhibit complex inter-dependencies.
Consider for example the following situation: A group of people
discuss about two correlated subjects, e.g., fish 
(in general) and salmon (in particular) as a part of diet. A person disliking fish also dislikes
salmon. If the influence process changes the individual's attitude
toward fish, say promoting fish as a healthy part of a diet, then such
a person may change her food preferences in favor of salmon as
well.

So far, only few pieces of work
\cite{nedic,fortunato,li2013consensus,tempo-science,Tempo-MD} have
tackled the dynamics of opinions on multiple, correlated subjects.  In
most of such pieces of work, opinions are represented as vector-valued
variables, evolving over a multidimensional space in which every axis
represents a different subject.
In particular, in \cite{nedic,fortunato,li2013consensus} the evolution of
opinions on different axes exhibit a weak dependency, i.e., two users interact only if
their Euclidean distance between opinions does not exceed a prefixed
threshold. The pieces of work in \cite{tempo-science, Tempo-MD}, instead,
propose a linear
multidimensional model that explicitly accounts
for the interdependence of opinions on various topics, and they provide conditions for
both stability and convergence \cite{Tempo-MD}.

{\em Our contribution and methodology.} In this paper, we move a step
forward with respect to the existing work. First, we generalize the model
in~\cite{tempo-science, Tempo-MD}, introducing a noise component,
which represents the individual endogenous process of opinion/belief
evolution.  Second, we enhance the model by accounting for possible
{\em repulsive} interactions due to adversion between individuals. Using
such a model and adopting a mean field approach holding for large
population of users, we characterize the evolution of opinions on
correlated subjects through a multidimensional Fokker-Planck
equation. We derive ergodicity conditions (i.e., conditions for the
existence of a unique stationary solution), and, under mild
assumptions, we obtain a {\em closed-form} expression for the stationary
distribution of individuals' opinions.  We remark that the stability
analysis in the presence of an adversarial individuals' attitude is much
 less obvious than in the traditional models (such as \cite{noi-TNSE}) where only attractive
forces were considered. Finally, we provide novel, efficient
numerical techniques to analyze both the steady state and the
transient solution of the Fokker-Planck (FP) equation, and show interesting
effects of  opinions dynamics and correlated topics in some relevant scenarios.

\subsection{Paper organization\label{sec:organization}}
The paper is organized as follows. In Sect.~\ref{sec:preliminaries} 
 we introduce the model and  we derive the  Fokker-Planck  equation based on mean-field approach.
 In Sect.~\ref{sec:FP} we  develop a methodology for the solution of the FP equation.  In Sect.~\ref{sec:stability} we analyse the stability of the system depending on the 
 different parameters. Sect.~\ref{sec:steady_state} presents an analysis of the steady-state regime. in Sect.~\ref{sec:summary} we summarize the mathematical tools we have used in the paper.
 Numerical results are reported in  Sect~\ref{sec:results}. Finally we draw some conclusions in Sect.~\ref{sec:conclusions}.

\subsection{Notation\label{sec:notation}}
Boldface uppercase and lowercase letters denote matrices ad column
vectors, respectively. $\Id_n$ is the identity matrix of size $n$ and
the transpose of the generic matrix $\Am$ is denoted by
$\Am\Tran$. The notation $\Am = \left\{a(i,j)\right\}$
  is sometimes used to define a matrix $\Am$ whose $(i,j)$-th element
  is $a_{i,j}$. Similarly, $\av = {\rm cat}\left\{\av(i)\right\}$
  indicates that the column vector $\av$ is obtained by concatenating
  the column vectors $\av_i$. The Laplace transform of the function
$f(x)$ is denoted by $\hat{f}(s)$. Finally, the symbol $\otimes$
denotes the Kronecker product.

\section{System model}\label{sec:preliminaries}
Consider a set of agents $\Uc$, with cardinality $U$, with agent $i$
exhibiting personality $P_i\in \Pc$. The agent's personality accounts
for her interests and habits, e.g., the social networks
to which she has subscribed or the forums in which she participates.
We consider that agents have opinions on $N$ different topics and that
an opinion formed on one subject is influenced by the opinions on some
of the other subjects, i.e., topics are interdependent
\cite{Tempo-MD,Tempo45,Tempo46}.  We define $\Cm$ as the coupling matrix, with
$c_{mn}$ expressing the entanglement of subject $m$ on subject $n$.
The opinions that agent $i\in \Uc$ has on the different subjects is
represented by a vector of size $N$, denoted by $\xv_i(t) \in \Xc^N$,
which evolves over continuous time, $t\in \RR_+$.
% $X_i(t)$ depends on the agent's personality, $P_i\in \Pc$, on the
% interaction with other agents, and on an endogenous random process
% accounting for the impact of individual experiences.
We define the prejudice vector $\uv(P_i)$ as the a-priori
$N$-dimensional belief that agent $i$ has on the different subjects;
also the prejudice depends on the agent's personality.  
%The
%opportunity that agents have to interact with each other is modeled
We represent through a graph the existence and the intensity of social
relationships between users, which depend on the personality of the
agents and on the similarity between the agents' beliefs.  The actual
influence that agents exert on each other then depends on their
opportunity to interact, as well as on their sensitivity to others' beliefs.

As a result, the evolution of agent $i$'s belief over time can be
represented as:
\begin{equation}
\xv_i(t\plus\dd t) = \xv_i(t)  +\Cm \dv_{\xv,i}(t)
\label{eq:x_i1}
\end{equation}
where $\xv_i(t)$ denotes the belief of agent $i$ on the $N$ topics at
the current time instant, $\dv_{\xv,i}(t)$ accounts for the variation of agent $i$  opinions in the time interval $[t,t+\dd t]$, and
$\Cm$ accounts for the influence of the opinion on one topic on the opinions on other topics. The quantity $\dv_{\xv,i}(t)$ is given by
\begin{eqnarray}
\label{eq:deltax}
  \dv_{\xv,i} &=& \frac{1\minus\alpha(P_i)}{U-1}\sum_{\substack{j\in \Uc \\ j \neq i}}   \underbrace{\zeta\left(P_i\mathord{,}P_j\right)\left[\xv_j(t)\minus \xv_i(t)\right] \dd t}_{(a)}\non
  &&\quad+ \underbrace{\alpha(P_i) \left[\uv(P_i)\mathord{-}\xv_i(t)\right] \dd t}_{(b)} + \underbrace{\sigma \dd \wv_i(t)}_{(c)} \,.
\end{eqnarray}
The meaning of the terms on the right hand side of the above
expression is as follows. 
\begin{itemize}
\item The first term $(a)$ represents the interaction of agent $i$ with
  all other agents in $\Uc$. In particular,
  \begin{itemize}
    \item $\alpha(P_i)\in (0,1]$ indicates how insensitive $i$ is to
      other agents' beliefs, which, as also discussed in
      \cite{sociology-community}, plays an important role in opinion
      dynamics. This parameter will also be referred to as the agent's
      level of stubbornness. When $\alpha(P_i)\to 1$, the agent becomes
      completely insensitive to others' beliefs (stubborn). Instead, as
      $\alpha(P_i)$ decreases, the agent is more inclined to
      accept others' beliefs and is less conditioned by her own
      prejudice. For brevity, in the following we denote
      $\bar{\alpha}(P_i) =1-\alpha(P_i)$;
    \item $\zeta(P_i,P_j)$ represents the presence and the
      strength of interactions between agents $i$ and $j$ (hereinafter
      also referred to as mutual influence).  It is a function of both
      agents' personality and defines the structure of the social
      graph~\cite{sociology-community}. Note that the interactions
      between agents do not depend on the proximity of their opinions,
      i.e., they are independent of $\xv_j(t)\minus \xv_i(t)$.  Also,
      whenever $\zeta(P_i,P_j) = 0$, the two agents do not influence
      each other, i.e., they never interact. Finally, it is fair
      to assume that each element of $\zeta(P_i,P_j)$ is upper bounded
      by a constant and is continuous with respect to its first and
      second arguments.
   \end{itemize}

\item The second term $(b)$ represents the tendency of an agent to retain her
  prejudice.  

\item The third term $(c)$ accounts for the endogenous process of the belief
  evolution within each agent. Such process is modeled as an
  i.i.d. standard Brownian motion with zero drift and scale parameter $\sigma^2$~\cite{Baccelli}.
\end{itemize}
We remark that $\xv_i(t+\dd t)$, i.e., the belief of agent $i$ at time $t+\dd t$,
depends on her personality $P_i$ and the current
agent's belief. In other words, the temporal evolution of agents' beliefs $
\{ \xv_i(t),\; i\in \Uc\}$ is Markovian over $\Yc^U$, where $\Yc =\Pc \times \Xc^N$ is an
$(N+1)$-dimensional continuous space.
Furthermore, in the following we assume that $\alpha(P_i)$, $\zeta(P_i,P_j)$, and each element of  $\uv(P_i)$  are
continuous with respect to their arguments. 
For ease of notation, we denote $\Zm(P_1,P_2) = \zeta(P_i,P_j)\Cm$;
thus, by replacing~\eqref{eq:deltax} in~\eqref{eq:x_i1}, the latter can be rewritten as
\begin{eqnarray}
\label{eq:x_i}
\xv_i(t\plus\dd t) &\hspace{-2ex}\mathord{=}&\hspace{-2ex} \xv_i(t)\plus\frac{1\minus\alpha(P_i)}{U-1}\hspace{-2ex}\sum_{\substack{j\in \Uc/\{i\}}}\hspace{-2ex}\Zm\left(P_i\mathord{,}P_j\right)\left[\xv_j(t)\minus \xv_i(t)\right]\dd t \non
&& +\alpha(P_i) \Cm\left[\uv(P_i)\mathord{-}\xv_i(t)\right] \dd t +\sigma\Cm \dd \wv_i(t) \,.
\end{eqnarray}

\subsection{From the discrete to  the continuous model}
We now extend the above model to the continuous case by using the mean-field theory.
We leverage on the procedure presented in~\cite{noi-TNSE} and apply it to the
multi-subject scenario. More in detail, we define the empirical
probability measure, $\rho^{(U)}(\dd p, \dd \xv, t)$ over $\Yc$ at
time $t$, as:
\begin{equation}
  \rho^{(U)} (\dd p, \dd \xv, t ) =\frac{1}{U}\sum_{i \in \Uc}
  \delta_{(P_i,\xv_i(t ))}(\dd p, \dd \xv)\,.
  \label{eq:rho_U}
\end{equation}
In the above expression, $\delta_{(P_i,\xv_i(t ))}(\dd p,\dd \xv)$ is
the Dirac measure centered at $(P_i,\xv_i(t ))$, i.e.,
$\delta_{(P_i,\xv_i(t ))}(\dd p,\dd \xv)$ represents the mass
probability associated with opinion $\xv_i(t )$ of agent $i$, which
has personality $P_i$. Note that in~\eqref{eq:rho_U} agents are seen
as particles in the continuous space $\Yc$, moving along the opinion
axis $\xv$.  As shown in \cite{noi-TNSE}, by applying the mean-field
theory~\cite{gartner1988mckean,dawson1983critical}, as $U\to \infty$,
$\rho^{(U)} (\dd p, \dd \xv, t)$ converges in law to the asymptotic
distribution $\rho(p, \xv, t)$, provided that
$\rho^{(U)} (\dd p, \dd \xv, 0)$ converges in law to $\rho(p,\xv,
0)$. Moreover, $\rho(p, \xv, t)$ can be obtained from the following
non-linear Fokker-Planck (FP)
equation~\cite{gartner1988mckean,dawson1983critical}:
%%%%%%%%%%%%%%%
\begin{align} \label{eq:fokker-planck}
\frac{\partial }{\partial t} \rho(p, \xv, t) = &-\sum_{n=1}^N \frac{\partial }{\partial x_n} \left[ \mu_n(p, \xv, t) \rho(p, \xv, t) \right] \nonumber \\ & + 
  \frac{1}{2}\sum_{m,n=1}^N D_{mn} \frac{\partial^2 }{\partial x_m \partial x_n}  \rho(p,\xv, t)
\end{align}
where $D_{mn}$ is the $(m,n)$-th entry of the diffusion tensor $\Dm = \sigma^2\Cm\Cm\Tran$, and $\mu_n(p, \xv, t)$ 
is defined as the component of the instantaneous
% average
speed along axis $x_n$ of a
generic agent whose personality and opinion at time $t$ are equal to $p$ and $\xv$. 
The  instantaneous speed is given by:
\begin{align}
\label{eq:mu_x2}
\muv(p,\xv ,t) =& \;\bar{\alpha}(p) \int_{\yv\in \Xc^N} \int_{\Pc} \Zm(p,q) (\yv-\xv)  \rho(q,\yv , t) \dd^N \yv \dd q  \nonumber \\ 
&\quad+ \alpha(p) \Cm [\uv(p) - \xv] \nonumber \\ 
=& \;\bar{\alpha}(p) \left[ \gammav(p,t) -  \Gammam(p) \xv \right] + \alpha(p) \Cm [\uv(p) - \xv] \nonumber \\ 
 = & \; -\Xim(p) \xv + \phiv(p,t)
%\label{eq:mu_s2}
\end{align}
where we defined
\begin{equation} \label{eq:Gammap}
\Gammam(p) \mathord{\triangleq}\hspace{-0.5ex}  \iint_{\yv,q} \hspace{-2ex}\Zm(p,q) \rho(q,\yv,t) \dd^N \yv \dd q \mathord{\stackrel{(a)}{=}}\hspace{-0.5ex} \int_q \hspace{-0.5ex}\Zm(p,q)   \rho_0(q) \dd q ,
\end{equation}
where in $(a)$ we wrote $\rho(q,\yv,t) =\rho(\yv,t |q)\rho_t(q)$ and
exploited the fact that by definition $\int_{\yv} \rho(\yv,t |q)\dd^N
\yv = 1 $. Since the distribution of the agents' personality at time
$t$, $\rho_t(q)$, does not depend on $t$, 
we have: $\rho_t(q)=\rho_0(q)$. Furthermore,
\begin{equation} \label{eq:J1}
\gammav(p, t) \triangleq  \iint_{\yv,q} \Zm(p,q)  \yv \rho(q,\yv , t) \dd^N \yv \dd q ,
\end{equation}
\begin{equation}  \label{eq:Xi}
\Xim(p) \triangleq \bar{\alpha}(p)\Gammam(p) + \alpha(p) \Cm ,
\end{equation}
\begin{equation} 
\phiv(p ,t) \triangleq \bar{\alpha}(p) \gammav(p, t) + \alpha(p) \Cm \uv(p) \label{eq:phi}
\end{equation}
and we considered a zero-drift Brownian motion process
$\wv(t)$.  
In the following, we analyze the system dynamics by solving the above
FP equation for $\rho(p, \xv, t)$ so as to obtain the
distribution of agents over $\Yc$.

\section{Solution of the Fokker-Planck (FP) equation} \label{sec:FP}

In this section, we solve the $N$-dimensional FP equation, which
describes an $N$-dimensional Ornstein-Uhlenbeck (OU)\cite{risken1996fokker} random process.
%whose Green function (impulse response) can be obtained as shown in Appendix~\ref{app:solution}.
%Using such Green function and c
Considering a general initial density $\rho(p,\xv, 0) = \rho_0(\xv|p)
\rho_0(p)$, we obtain the solution of the FP equation $\rho(p,\xv, t)$
as shown in Appendix~\ref{app:solution} in the Supplemental Material:
\begin{equation}
\rho(p,\xv, t) %&=&  \int_{\yv} \rho(p,\xv ,t|\yv ) \rho(p,\yv, 0)   \dd^N \yv \\
               \mathord{=} \rho_0(p) \int_{\yv}\hspace{-1ex} \Gc\left(\xv\mathord{,} \mv(p\mathord{,}\yv\mathord{,}t)\mathord{,} \Sigmam (p\mathord{,}t) \right)
               % \frac{\exp\left\{ -\frac1{2} \left(\xv - \mv(p,\yv,t)\right) \Tran \Sigmam^{-1}(p ,t) \left(\xv - \mv(p,\yv,t)\right) \right\}}{ (2 \pi)^{N/2} |\Sigmam(p ,t) |^{1/2}}
                     \rho_0(\yv|p)\dd^N \yv \label{eq:solution}
\end{equation}
where $\Gc\left(\xv, \mv(p,\yv,t), \Sigmam (p ,t) \right)$ is the pdf of the Gaussian multivariate distribution with
covariance
\begin{equation}\label{eq:covariance}
\Sigmam(p ,t) \triangleq \int_0^{t} \ee^{-\Xim(p) \tau} \Dm \ee^{-\Xim(p)\Tran \tau} \dd\tau
\end{equation}
and mean 
\begin{eqnarray}
  \mv(p,\yv,t) &\hspace{-2ex}\triangleq&\hspace{-2ex}\ee^{-\Xim(p) t} \left[ \yv+\int_0^{t} \ee^{\Xim(p) \tau} \phiv(p,\tau)\dd\tau \right] \non
&\hspace{-2ex}\stackrel{(a)}{=}&\hspace{-2ex} \ee^{-\Xim(p) t} \yv + \alpha(p)\Xim^{-1}(p) \left(\Id_N\minus\ee^{-\Xim(p) t} \right) \Cm \uv(p)\nonumber \\
&\hspace{-2ex} &+ \bar{\alpha}(p) \int_0^{t} \ee^{-\Xim(p) (t-\tau)} \gammav(p,\tau) \dd\tau \label{eq:mv_x}
\end{eqnarray}
where in (a) we used the definition of $\phiv(p,\tau)$ provided in~\eqref{eq:phi}.
However, notice that $\mv(p,\yv,t)$ in \eqref{eq:mv_x} is a function of 
$\gammav(p,t)$, which in turn is a function of $\rho(p,\xv, t)$ (see
\eqref{eq:J1}). As such, we have to impose a self-consistency
condition; precisely, replacing~\eqref{eq:solution} in~\eqref{eq:J1}, we obtain
\begin{eqnarray}
  \gammav(p,t)
  &\hspace{-2ex}=&\hspace{-2ex} \int_{\yv} \int_q  \Zm(p,q)  \yv \rho(q,\yv , t) \dd^N \yv \dd q \non
  &\hspace{-2ex}=&\hspace{-2ex} \int_{\yv} \int_q  \Zm(p,q)  \yv \rho_0(q)   \int_{\zv} \Gc\left(\yv, \mv(q,\zv,t), \Sigmam(q ,t)\right) \non
  &\hspace{-2ex} &\hspace{-2ex}\qquad\cdot  \rho_0(\zv|q)\dd^N \zv \dd^N \yv \dd q \non
  &\hspace{-2ex}=&\hspace{-2ex} \int_q \Zm(p,q)  \rho_0(q) \int_{\zv} \mv(q,\zv,t) \rho_0(\zv|q)\dd^N \zv \dd q \non
  &\hspace{-2ex}=&\hspace{-2ex} \gammav_0(p,t) + \gammav_1(p,t) +\int_q \Zm(p,q)  \rho_0(q)\bar{\alpha}(q) \non
  &\hspace{-2ex}&\hspace{-2ex}\qquad\cdot\int_0^{t} \ee^{-\Xim(q) (t-\tau)} \gammav(q,\tau)\dd\tau \dd q
\label{eq:fixed_point}
\end{eqnarray}

\noindent where we  used~\eqref{eq:mv_x} and defined for brevity:
\begin{equation}
  \gammav_0(p,t) \triangleq \int_q \Zm(p,q)  \rho_0(q) \ee^{-\Xim(q) t} \int_{\yv} \yv \rho_0(\yv|q)\dd^N \yv \dd q
  \label{eq:gamma0}
\end{equation}
\begin{eqnarray}
  \gammav_1(p,t) & \triangleq & \int_q \Zm(p,q)  \rho_0(q) \alpha(q) \Xim^{-1}(q)\non
                                &&\qquad \cdot \left(\Id_N\minus\ee^{-\Xim(q) t} \right) \Cm \uv(q)  \dd q
  \label{eq:gamma1}
\end{eqnarray}
Interestingly, \eqref{eq:fixed_point} is a linear Volterra equation of the
second kind \cite{Volterra}. 
We can take over time the Laplace transform of \eqref{eq:fixed_point} and get
\begin{eqnarray} \label{eq:fixed_point_transform}
\widehat{\gammav}(p,s) &\hspace{-1ex}=& \hspace{-1ex} \widehat{\gammav}_0(p,s) + \widehat{\gammav}_1(p,s) \nonumber \\
  &\hspace{-3ex} & \hspace{-3ex} +\int_q\hspace{-0.5ex} \Zm(p,q)  \rho_0(q) \bar{\alpha}(q) \Xm^{-1}(s,q)\widehat{\gammav}(q, s) \dd q \non
\end{eqnarray}
where $\Xm(s,q) = s \Id_N + \Xim(q)$
\begin{equation}
\widehat{\gammav}_0(p,s) = \hspace{-1ex} \int_q\hspace{-0.5ex}  \Zm(p,q) \rho_0(q)  \Xm^{-1}(s,q)\hspace{-1ex} \int_{\yv} \yv \rho_0(\yv|q)\dd^N \yv \dd q\label{eq:fixed_point_transform_0}
\end{equation}
\begin{equation}
\widehat{\gammav}_1(p,s) =\hspace{-1ex} \int_q  \Zm(p,q) \rho_0(q)  \alpha(q) s^{-1} \Xm^{-1}(s,q) \Cm \uv(q)  \dd q \label{eq:fixed_point_transform_1}
\end{equation}

Eq.\,\eqref{eq:fixed_point_transform} is a non-homogeneous integral equation, 
whose solution is unique if and only if the associated homogeneous equation has no nonzero solutions. If the solution is unique, then it gives the solution of the FP through \eqref{eq:solution}. %Otherwise, the solution of the FP will lie somewhere in a space with the same dimension as the solution space of the homogeneous 
%integral equation (independently from the initial conditions). 

\textbf{Remark}: 
By restricting the definition of $\gammav(p,t)$ over an arbitrary
compact domain $\mathcal {P}\times [0,T]$,
\eqref{eq:fixed_point} can be rewritten in an operational  form as 
$(\Ic-\Ac)[\gammav(p,t)]=\gammav_0(p,t) + \gammav_1(p,t)$ where the
operator $\Ac[\gammav(p,t)]=\int_q \Zm(p,q)  \rho_0(q)
\bar{\alpha}(q) \int_0^{t} \ee^{-\Xim(q) (t-\tau)} \gammav(q,\tau)
\dd\tau \dd q $. Note that, as an immediate consequence of  {the structure of   Volterra equations over compact domains}, we have
$\|\Ac^n\| < 1$ for sufficiently large $n$ \cite[cap.\,2. pp.\,50--51]{Kolmogorov}. Thus, 
$(\Ic-\Ac)[\cdot]$ is invertible (i.e., the solution is unique) and an expression for
$\gammav(p,t)$ can be obtained as
$\gammav(p,t)= (\Ic-\Ac)^{-1}[\left(\gammav_0(p,t) + \gammav_1(p,t)\right)]= \sum_n \Ac^n [\left(\gammav_0(p,t) + \gammav_1(p,t)\right)]$.
\begin{comment}
Now a sufficient condition  for operator $ (\Ic-\Ac)^{-1})$ to be bounded  in norm is that $\|\mathcal A\|<1$. (Indeed, under the assumption that   $\|\mathcal A\|<1$ it holds:
$\|(I- \mathcal A^{-1}) \|<\frac{1}{1-\| \mathcal A\|}$.
Therefore a sufficient  condition  for the  boundedness of $\gamma(p,t)$ is $\|\mathcal A\|_\infty<1$.

By direct inspection it can be easily shown that $\|\mathcal A\|_\infty<1$ whenever  $Z(p,q)\ge 0$ and $\inf_p \alpha(p)>0$.
Indeed:
\[
 \|\mathcal A\|_\infty\le  \int_q \left| {\Zm(p,q)  \rho_0(q) (1-\alpha(q))  } \right| \int_0^{t} \ee^{-\Xim(q) (t-\tau)}  d\tau \dd q\le  \int_q \left| {\Zm(p,q)  \rho_0(q) (1-\alpha(q))\Gamma(q) \Xim^{-1}(q)} \right| \dd q
\]
with
\[
  \int_q \left| {\Zm(p,q)  \rho_0(q) (1-\alpha(q))  \Xim^{-1}(q)} \right| \dd q= \int_q  {\Zm(p,q)  \rho_0(q) (1-\alpha(q)) \Xim^{-1}(q)}  \dd q\le \sup_q \frac{ (1-\alpha(q)) \Gamma(q) }{\alpha(q) + (1-\alpha(q))\Gamma(q) }:=\beta<1
\]
Then  over every compact domain   $\mathcal {P}\times [0,T]$, we have
$\|\theta(p,t)\|_\infty< \| \frac{\gammav_0(p,t) + \gammav_1(p,t)}{\Gamma(p)} \|_\infty \frac{1}{1-\beta}$.
In other words,   given that  $\| \frac{\gammav_0(p,t) + \gammav_1(p,t)}{\Gamma(p)} \|_\infty$ is uniformly bounded with respect to $T$ 
 also   $\|\theta(p,t)\|_\infty$  is uniformly bounded with respect to $T$.
\end{comment}

Provided that the integral equation~\eqref{eq:fixed_point_transform} admits a unique solution,
in general  it is still rather challenging to explicitly find it. In the
following, we will particularize our analysis  to two cases in which it is
possible to find an explicit analytical expression for  such as
solution. % of \eqref{eq:fixed_point_transform}.

%\color{red} DA FARE scrivere un cappello in cui si descrive cosa si fa in questa sezione.}

\subsection{The case of $\zeta(p,q)$ in product form} \label{sec:easy}

We recall that $\Zm(p,q)=\zeta(p,q)\Cm$.  If
$\zeta(p,q) = \zeta_1(p) \zeta_2(q)$, then from~\eqref{eq:J1} we have
\begin{eqnarray}
  \gammav(p, t)
  &=& \zeta_1(p) \int_{\yv} \int_q  \zeta_2(q)\Cm\yv \rho(q,\yv , t) \dd^N \yv \dd q \,. \non
  &=& \zeta_1(p)\bv(t) \label{eq:gamma_delta}
\end{eqnarray}
Taking the Laplace transform of~\eqref{eq:gamma_delta}, we obtain
$\widehat{\gammav}(p, s)=\zeta_1(p)\widehat{\bv}(s)$.  Using the
latter expression in~\eqref{eq:fixed_point_transform} and
recalling~\eqref{eq:fixed_point_transform_0} and
\eqref{eq:fixed_point_transform_1}, we get
\begin{eqnarray} \label{eq:product_fixed_point_transform}
  \widehat{\gammav}(p,s)
  &=&  \widehat{\gammav}_0(p,s) + \widehat{\gammav}_1(p,s) \non
  &&\hspace{-5ex} +\zeta_1(p)\int_q \zeta_2(q) \Cm \rho_0(q) \bar{\alpha}(q) \Xm^{-1}(s,q)\widehat{\gammav}(q, s) \dd q  \non
  &=&  \zeta_1(p)\left(\widehat{\bv}_0(s)+ \widehat{\bv}_1(s)\right)\non
  && \hspace{-5ex}+\zeta_1(p)\int_q\rho_0(q) \bar{\alpha}(q) \zeta_2(q) \Cm  \Xm^{-1}(s,q)\zeta_1(q) \dd q \widehat{\bv}(s) \non
\end{eqnarray}
under the assumption that  $\Xm(s,q) = s \Id_N + \Xim(q)$ is
invertible\footnote{Note that invertibility is granted for $\mathrm{Re}(s)>\sup_q \|\Xim(q)\|$}. In~\eqref{eq:product_fixed_point_transform}
\begin{equation}
\widehat{\bv}_0(s) = \int_q  \zeta_2(q)\Cm\rho_0(q) \Xm^{-1}(s,q)\hspace{-1ex} \int_{\yv} \yv \rho_0(\yv|q)\dd^N \yv \dd q\label{eq:product_fixed_point_transform_0}
\end{equation}
\begin{equation}
\widehat{\bv}_1(s) = \int_q  \zeta_2(q)\Cm \rho_0(q)  \alpha(q) s^{-1}\Xm^{-1}(s,q) \Cm \uv(q)  \dd
q \label{eq:product_fixed_point_transform_1} \,.
\end{equation}
Note that we can also write 
\begin{align} \label{eq:fixed_point_transform_sep}
\widehat{\bv}(s) =& \left( \int_q   \rho_0(q) \bar{\alpha}(q) \zeta_2(q) \Cm  \Xm^{-1}(s,q) \zeta_1(q) \dd q \right) \widehat{\bv}(s)\non
  &+\widehat{\bv}_0(s) + \widehat{\bv}_1(s) 
\end{align}
from which, under the assumption that matrix
\[ \Tm(s)= \Id_N -\int_q \rho_0(q) \bar{\alpha}(q) \zeta_2(q) \Cm \Xm^{-1}(s,q) \zeta_1(q) \dd q\]
is non singular, we obtain:
\[ \widehat{\bv}(s) =  \Tm(s)^{-1}\left( \widehat{\bv}_0(s) + \widehat{\bv}_1(s)\right)\,.\]
We now have an explicit solution (in the transformed domain) for
$\bv(t)$. Indeed,  for $\mathrm{Re}(s)$ sufficiently large, matrix
$\Tm(s)$ is non singular. Once
$\bv(t)$ is obtained, we can compute $\gammav(p,t)$
through~\eqref{eq:gamma_delta}, then $\mv(p,\yv,t)$
in~\eqref{eq:mv_x}, and, finally, 
our opinion density $\rho(p,\xv,t)$ through~\eqref{eq:solution}.

%{\color{red} Questa parte andrebbe spostata nella sezione stazionaria
%
%  
%Under this condition, from $\widehat{\bv}(s)$ we obtain the following stationary solution for $\gammav(p,t)$ if $\Xim(p)$ is Hurwitz-stable
%\begin{equation}
%\gammav(p,+\infty) = \zeta_1(p)\left(\Id_N -\int_q   \rho_0(q) \bar{\alpha}(q) \zeta_2(q) \Cm  \Xim(q)^{-1} \zeta_1(q) \dd q \right)^{-1} \int_q \rho_0(q)  \alpha(q) \zeta_2(q)  \Xim^{-1}(q)  \Cm^2 \uv(q)  \dd q 
%\end{equation}
%}
\subsection{Discrete personality distribution}
Suppose that the personality distribution is discrete with $M$ probability masses; then we write the opinion distribution at $t=0$ as
\begin{equation}\label{eq:rho0}
\rho_0(p) = \sum_{i=1}^M r_i\delta(p-p_i).
\end{equation}
Consequently, the fixed-point equation \eqref{eq:fixed_point_transform} becomes
\begin{eqnarray} \label{eq:fixed_point_transform_disc}
  \widehat{\gammav}(p_i,s)&=& \sum_{k=1}^M \Zm(p_i,p_k) r_k \bar{\alpha}(p_k) \Xm^{-1}(s,p_k)\widehat{\gammav}(p_k,s)\non
                        &&\qquad +\widehat{\gammav}_{0}(p_i,s) + \widehat{\gammav}_{1}(p_i,s)\,.
\end{eqnarray}
In~\eqref{eq:fixed_point_transform_disc}, we defined
\begin{equation}  \label{eq:gamma0_disc}
\widehat{\gammav}_0(p_i,s) =\sum_{k=1}^M \Zm(p_i,p_k) r_k \Xm^{-1}(s,p_k)\xv_{0}(p_k)
\end{equation}
where $\xv_{0,k}$ is the average opinion at time $t=0$ corresponding to personality $p_k$ and has been obtained by observing that the term $\rho_0(q)\int_{\yv}\yv \rho_0(\yv|q)\dd^N \yv$ in~\eqref{eq:fixed_point_transform_0} is equal to $\mv(q,\xv,0)=\xv_0(q)$.
Also, we defined: 
\begin{equation} \label{eq:gamma1_disc}
\widehat{\gammav}_1(p_i,s) =\sum_{k=1}^M \frac{\Zm(p_i,p_k) r_k \alpha(p_k)}{s} \Xm^{-1}(s,p_k)\Cm \uv(p_k) \,.
\end{equation}
Next, recalling the notation defined in~Section~\ref{sec:notation} we define
  $\underline{\widehat{\gammav}}(s) = {\rm cat}\{\widehat{\gammav}(p_i,s)\}$,
  $\underline{\widehat{\gammav}}_0(s)={\rm cat}\{\widehat{\gammav}_0(p_i,s)\}$,
  $\underline{\widehat{\gammav}}_1(s) ={\rm cat}\{\widehat{\gammav}_{1}(p_i,s)\}$,
  $\underline{\xv}_0 = {\rm cat}\{\xv_0(q_i)\}$,
  $\underline{\uv} = {\rm cat}\{\uv(q_i)\}$,
  $\underline{\Xim} = \diag \left(\Xim(p_1), \dots, \Xim(p_M) \right)$,  
  $\Pm_0 = \diag \left(r_1, \dots, r_M \right) \otimes \Id_N$, $\Pm_1 = \diag \left(r_1 \alpha(p_1), \dots, r_M \alpha(p_M) \right) \otimes \Id_N$,
  $\Pm_2 = \diag \left(r_1 \bar{\alpha}(p_1), \dots, r_M\bar{\alpha}(p_M) \right) \otimes \Id_N$, and
\begin{equation}
  \underline{\Zm} = \left\{\Zm(p_i,p_j)\right\} = \left\{  \zeta(p_i,p_j)\right\}\otimes \Cm\,.
\end{equation} 
Then, we can write \eqref{eq:fixed_point_transform_disc} as
\begin{equation} \label{eq:fixed_point_transform_mat}
\underline{\widehat{\gammav}}(s) = \underline{\widehat{\gammav}}_0(s) + \underline{\widehat{\gammav}}_1(s) + \underline{\Zm} \left(s \Id_{MN} + \underline{\Xim} \right)^{-1}\Pm_2  \underline{\widehat{\gammav}}(s)
\end{equation} 
while \eqref{eq:gamma0_disc} and \eqref{eq:gamma1_disc} become
\begin{equation} \label{eq:gamma0_mat}
\underline{\widehat{\gammav}}_0(s) = \underline{\Zm} \left(s \Id_{MN} + \underline{\Xim} \right)^{-1}\Pm_0  \underline{\xv}_0
\end{equation} 
and
\begin{equation} \label{eq:gamma1_mat}
\underline{\widehat{\gammav}}_1(s) = \underline{\Zm} \left(s \Id_{MN} + \underline{\Xim} \right)^{-1}s^{-1} \Pm_1  \left(\Id_M \otimes \Cm \right) \underline{\uv}.
\end{equation}
Let $\underline{\Xm}(s)=s\Id_{MN} + \underline{\Xim}$. Then we can solve explicitly the fixed-point equation of \eqref{eq:fixed_point_transform_mat} (in the transform domain) as
\begin{eqnarray}
\underline{\widehat{\gammav}}(s) &=& \left(\Id_{MN}\minus \underline{\Zm} \underline{\Xm}(s)^{-1}\Pm_2 \right)^{-1} \left( \underline{\widehat{\gammav}}_0(s) \plus\underline{\widehat{\gammav}}_1(s) \right)\non
                                 & = & \left(\Id_{MN}\minus \underline{\Zm} \underline{\Xm}(s)^{-1}\Pm_2 \right)^{-1} \underline{\Zm} \underline{\Xm}(s)^{-1}\non
                                       && \qquad \cdot\left( \Pm_0  \underline{\xv}_0\plus s^{-1} \Pm_1  \left(\Id_M \otimes \Cm \right)\underline{\uv}\right) \non
& = & \left(\underline{\Xm}(s)\underline{\Zm}^{-1} \minus \Pm_2 \right)^{-1} \left( \Pm_0  \underline{\xv}_0 \plus s^{-1} \Pm_1  \left(\Id_M \otimes \Cm \right)\underline{\uv}\right) \non
& = & \underline{\Zm} \left(\underline{\Xm}(s)\minus \Pm_2 \underline{\Zm} \right)^{-1} \left( \Pm_0  \underline{\xv}_0 \plus s^{-1} \Pm_1  \left(\Id_M \otimes \Cm \right)\underline{\uv} \right)\,.\non
\end{eqnarray}

Now, let 
\begin{equation}\label{eq:Psim}
\Psim = \underline{\Xim}  - \Pm_2 \underline{\Zm} \,. 
\end{equation}
Provided that $ \Psim$ is invertible, the above solution can be inverse-transformed to get the solution in the time domain as
\begin{equation} \label{eq:gamma_t}
\underline{\gammav}(t) = \underline{\Zm} \left(\ee^{-\Psim t} \Pm_0  \underline{\xv}_0\plus \Psim^{-1} \left(\Id_{MN}\minus\ee^{-\Psim t} \right) \Pm_1 \left(\Id_M \otimes \Cm \right) \underline{\uv} \right). 
\end{equation}
Then $\rho(p,\xv,t)$ can be obtained as described in the previous
section.

Finally, the following theorem complements the above result:
\begin{theorem}\label{theo1}
	i) For any finite $t$ the Volterra equation %\eqref{eq:beta_Fredholm} 
	\eqref{eq:fixed_point} admits a unique
	solution which is Lipschitz-continuous.  ii) The solution $\gamma(p,t)$ of the Volterra
	equation	\eqref{eq:fixed_point} , under any distribution
	$\rho_0(p)$, which is continuous at every point in $p$, is the
	uniform limit of solutions $\gamma_n(p,t)$, obtained by replacing
	distribution $\rho_0(p)$ with its discrete approximation $\rho_n(p)$
	whose mesh-size is $\frac{1}{n}$.
\end{theorem}
The proof of this theorem is given following exactly  the same lines of Appendix~\ref{app:C} (which proves a similar statement under  steady state conditions).

%{\color{red} da mettere nella parte stazionaria
%If $\Psim$ is Hurwitz-stable, the limit solution for $t \rightarrow \infty$ is given by
%\begin{equation}
%\underline{\gammav}(\infty) = \underline{\Zm}  \Psim^{-1}  \Pm_1 \Cm \underline{\uv} . 
%\end{equation}
%}
\section{Stability analysis} \label{sec:stability}
Let us define the system as \emph{stable} if, for every initial
condition, the opinion distribution converges to the stationary
solution for $t \rightarrow \infty$. 
In this section, we first focus on the case where the personality distribution is
discrete, then we generalize the result to the the case of continuous personality distributions.

\subsection{Discrete personality distribution}
Let us state the stability conditions as follows:
\begin{itemize}
\item $\underline{\Xim}$ is Hurwitz-stable (i.e., all the eigenvalues of $\underline{\Xim}$ have positive real part).
\item $\Psim$ is Hurwitz-stable.  
\end{itemize}
Indeed, recalling the expression of the covariance in~\eqref{eq:covariance}, its
value remains limited if the first condition is met, while, looking
at~\eqref{eq:mv_x} and~\eqref{eq:gamma_t}, the mean of the
distribution remains limited if both the above conditions are satisfied.
In particular, when the first condition is not met, there are some
personalities for which the opinion distribution scatters about along
some directions. We call this phenomenon \emph{type-I instability}.
Instead, if the first condition is met and the second one is not, for all
personalities the opinion covariance remains limited but there are some
personalities for which the mean opinion value drifts to
infinity. We will refer to this case as \emph{type-II instability}.

Below we elaborate on the conditions that are needed to ensure the system stability.

\subsubsection{Condition to avoid Type-I instability}
Let $\lambda_k(\Am) = \lambda^R_k(\Am) + j \lambda_k^I(\Am)$ be the
$k$-th eigenvalue of matrix $\Am$. Recall that $\Xim(p_i)=\bar{\alpha}(p_i)\Gammam(p_i)+\alpha(p_i)\Cm$,
thus the necessary and sufficient condition to avoid type-I instability is
given by
\[ \min_k \lambda^R_k(\Xim(p_i)) > 0, \quad \forall i\,.\] 
Since $\Gammam(p_i) = \Cm \sum_h \zeta(p_i,p_h) \rho_0(p_h)$, we can write
\[ \Xim(p_i)=\left[\bar{\alpha}(p_i)\sum_h \zeta(p_i,p_h) \rho_0(p_h)+\alpha(p_i)\right]\Cm\,.  \]
It follows that the stability condition becomes
\[ \left[\bar{\alpha}(p_i)\sum_h \zeta(p_i,p_h) \rho_0(p_h)+\alpha(p_i)\right] \min_k \lambda^R_k(\Cm)  > 0\,. \]
Assuming that $\bar{\alpha}(p_i)\sum_h \zeta(p_i,p_h) \rho_0(p_h)+\alpha(p_i)>0$ for every $i$, then
stability is ensured when
\[  \min_k \lambda^R_k(\Cm)  >0 \]
%Otherwise if $\sum_h \zeta(p_i,p_h) \rho_0(p_h)+\alpha(p_i)< -\alpha(p_i)/\bar{\alpha}(p_i)$ for every $i$
%then the stability condition becomes 
%\[ \max_k \lambda^R_k(\Cm) <0 \]
The above expression highlights
that if the opinion dynamics along every topic are stable, introducing
a Hurwitz stable matrix $\Cm$ preserves stability.

\subsubsection{Condition to avoid Type-II instability} 
The necessary and sufficient condition to avoid type-II instability is
given by
\[ \min_k \lambda^R_k(\Psim(p_i)) > 0, \quad \forall i\,.\] 
Let us focus on the case where
$\alpha(p) \equiv \alpha$. By recalling~\eqref{eq:Psim}, we write
\begin{eqnarray}
  \Psim &=& \underline{\Xim} - \Pm_2\underline{\Zm} \non
  &=& \bar{\alpha}\underline{\Gammam}+\alpha\Id_{M}\otimes \Cm -\bar{\alpha}\Pm_0\underline{\Zm} \non
      &=&  \alpha \Id_{M}\otimes \Cm + \bar{\alpha} \left( \underline{\Gammam} - \Pm_0 \underline{\Zm} \right) \label{eq:Psim2}
\end{eqnarray}
where
$\underline{\Gammam} = \diag \left(\Gammam(p_1), \dots, \Gammam(p_M)
\right)$, with $\Gammam(p_i)$ obtained by
discretizing~\eqref{eq:Gammap}. Then 
we define
$\underline{\Gammam} - \Pm_0 \underline{\Zm} \triangleq \Thetam
\otimes \Cm$, where $\Thetam$ is the $M \times M$ matrix whose
$(i,j)$-th entry is given by
\[
\theta_{ij} = \left\{
\begin{array}{rl}
\sum_{k \neq i} \zeta(p_i,p_k) r_k,  & i=j \\
- \zeta(p_i,p_j) r_i , & i\neq j \,.
\end{array} \right.
\]  
%Consequently, the stability condition becomes
%\[
%\min_i \lambda^R_i(\underline{\Gammam} - \Pm_0 \underline{\Zm}) > -\frac{\alpha}{1-\alpha}
% \]
As a consequence, considering that $\lambda_{(j-1)N +i}(\Thetam\otimes
\Cm)= \lambda_{i}(\Thetam)\lambda_j(\Cm)$ and using~\eqref{eq:Psim2},
the condition for stability reads as follows:
\[
\Re\{\lambda_i(\Thetam) \lambda_j(\Cm)\} > -\frac{\alpha}{1-\alpha}
\Re\{\lambda_j(\Cm)\}\,\,\,\,\,\, \forall i,j \,.
\] 

{\bf Remark.} In the scalar case ($N=1$), the following propositions
hold. % (the proofs are reported in Appendix~\ref{app:proofs_prop}).
\begin{proposition} \label{prop:stab_1}
If $\zeta(p_i,p_j) \geq 0$ for all $i,j$, $\Psim$ is Hurwitz-stable.
\end{proposition}

\pf If $\zeta(p_i,p_j) \geq 0$ for all $i,j$, $\Psim$ is a matrix whose off-diagonal elements are nonpositive and
we can apply Theorem 1 of \cite{Plemmons}. Precisely the Hurwitz stability of
$\Psim$ (condition J29 in Theorem 1 of \cite{Plemmons}) is implied by
the fact that the row sums of $\Psim$ are all positive (condition K35
applied with diagonal matrix $\Dm=\Id$). Since the row sums of $\Psim$
are all equal to $\alpha$, the proposition follows.\qedsymb

\begin{proposition} \label{prop:stab_2}
$\Psim$ is Hurwitz-stable if $\min_{i,j}\zeta(p_i,p_j) < 0$ and, for every $i$
\beq \label{eq:cond_stab}
\sum_{i \neq j} \left( \zeta(p_i,p_j) r_j- |\zeta(p_i, p_j)| r_i
\right)  > -\frac{\alpha}{1-\alpha} \,.
\eeq
\end{proposition}

\pf If $\min_{i,j}\zeta(p_i,p_j) < 0$, the Hurwitz stability of $\Psim$ (condition J29 in
Theorem 1 of \cite{Plemmons}) is implied by condition N39 applied with
diagonal matrix $\Dm=\Id$, which is expressed in
\eqref{eq:cond_stab}. 
\qedsymb

Note that the above propositions still hold in the case of continuous personalities.
% Although the condition in \eqref{eq:cond_stab} is not necessary, it
% turns out that for some cases it is tight, as in the following
% example.
We now present an example on the stability conditions for a simple case as described below.

\example{1}{Consider $N=1$ and that
there are $M$ personalities $p_i =\frac{(2i-1)}{M}-1$, $i=1,\dots,M$
($M$ even), with $r_i= \frac1{M}$ and
\[
\zeta(p_i,p_j) = \left\{
\begin{array}{rl}
\zeta_1, & p_i p_j > 0 \\
- \zeta_2 , & p_i p_j < 0 
\end{array} \right.
\]
with $\zeta_1,\zeta_2$ being arbitrary positive values. Thus,
$\Xim_i$ is a scalar equal to
$\alpha + \bar{\alpha} \frac{\zeta_1-\zeta_2}{2}$ for all
$i$. Moreover,
\[
\underline{\Zm} = \left[
\begin{array}{cc}
\zeta_1 & - \zeta_2 \\
- \zeta_2  & \zeta_1
\end{array} \right] \otimes \mathbf{1}_{M/2}
\]
where $\mathbf{1}_{n}$ is a size-$n$ square matrix with all entries equal to 1. Thus, 
\[
\lambda_{i}(\underline{\Zm}) = \left\{
\begin{array}{rl}
0, & 1 \leq i \leq M-2 \\
M \frac{\zeta_1-\zeta_2}{2} , & i = M-1 \\
M \frac{\zeta_1+\zeta_2}{2} , & i = M \,.
\end{array} \right.
\]
Since $\Psim = \left(\alpha + \bar{\alpha} \frac{\zeta_1-\zeta_2}{2}\right)\Id_M - \frac{\bar{\alpha}}{M} \underline{\Zm}$, it follows that
\[
\lambda_{i}(\Psim) = \left\{
\begin{array}{rl}
\alpha + \bar{\alpha} \frac{\zeta_1-\zeta_2}{2}, & 1 \leq i \leq M-2 \\
\alpha , & i = M-1 \\
\alpha - \bar{\alpha} \zeta_2 , & i = M \,.
\end{array} \right.
\]
We then have type-I instability if $\zeta_2 \geq \zeta_1 + 2 \frac{\alpha}{1-\alpha}$ and type-II
instability if 
$ \frac{\alpha}{1-\alpha} \leq \zeta_2 < \zeta_1 + 2
\frac{\alpha}{1-\alpha}$. Conversely, the system is stable iff
$\zeta_2 < \frac{\alpha}{1-\alpha}$. It is easy to see that this is 
equivalent to condition \eqref{eq:cond_stab}, which in this case is
necessary and sufficient. \qedsymb
}

\subsection{Continuous personality distribution}

We consider the case of a continuous personality distribution as the
limit case of a family of discrete distributions with increasingly
small discretization steps, $\Delta p$.
Then  similarly to what done in the previous section, we assume that
the following inequality holds:
\begin{align} \label{eq:stab_cond_cont}
   & \lim_{\Delta p \to 0}\left(\sum_{j \neq i} \zeta(p_i,p_j) \rho_{0}(p_j) \Delta p - \sum_{j \neq i} |\zeta(p_i,p_j)| \rho_{0}(p_i) \Delta p\right) \non
  &\quad= \int \zeta(p,q) \rho_0(q) \dd q - \int |\zeta(p,q)| \dd q \rho_0(p) \non
& \quad > -\frac{\alpha(p)}{1-\alpha(p)} \,.
\end{align} 
We now prove that the above is the stability condition for
the continuous case, provided that some technical conditions, specified below,  are met.
Suppose that \eqref{eq:stab_cond_cont} is true for our
continuous-personality system. Then, for $\Delta p$ sufficiently
small, \eqref{eq:cond_stab} is also satisfied for the discretized
system, so that the corresponding $\Psim(\Delta p)$ is Hurwitz-stable. main-field 

Next we assume that  for $\Delta p$ sufficiently small:
\begin{itemize}
\item $\Psim(\Delta p)$  is uniformly Hurwitz-stable, i.e., the minimum real part of its eigenvalues is bounded away from zero by an amount $\epsilon$ and
\item $\Psim^{-1}(\Delta p)$ has a uniformly bounded norm, i.e., $\| \Psim^{-1}(\Delta p)\| < K,$  $\forall \Delta p$ sufficiently small and  $K<\infty$~\footnote{This last condition is implied by the previous one when $\Psim(\Delta p)$ is diagonalizable. }.  
\end{itemize}
It follows that, for $\Delta p$ sufficiently small, the fixed-point
solution in \eqref{eq:gamma_t} is uniformly bounded from
above. As $\Delta p \rightarrow 0$, the fixed-point solution for the
continuous-personality system is  uniformly bounded from
above (by Theorem \ref{theo1}  and norm continuity) for every finite $t$. Hence, the system
is stable.

\section{Steady-state analysis\label{sec:steady_state}} 
Under previous stability/ergodicity conditions, it is interesting to
analyse the limiting solution for $t\to \infty$, which can be obtained
as solution of the associated steady state FP equation.
In the most general case, we can rewrite~\eqref{eq:fokker-planck} disregarding the dependence on $t$, as the following multi-dimensional FP equation:
 \begin{eqnarray} \label{eq:fokker-planck-multi}
\sum_{n=1}^N \frac{\partial }{\partial x_n} \left( \mu_n(p, \xv) \rho(p, \xv) \right)\mathord{=} 
  \sum_{m,n=1}^N \frac{D_{mn}}{2} \frac{\partial^2\rho(p,\xv) }{\partial x_m \partial x_n}  
\end{eqnarray}
where $\rho(p,\xv) =\lim_{t \to \infty}\rho(p,\xv,t) $ and, 
according to the second line in~\eqref{eq:mu_x2}, $\muv$ is given by 
\begin{eqnarray} \label{eq:mu_steady_state}
  \muv(p,\xv)
  &=& \bar{\alpha}(p) \left[ \gammav(p) -  \Gammam(p) \xv \right] + \alpha(p) \Cm [\uv(p) - \xv] \non
  &=&  \bar{\alpha}(p)\Cm( \betav(p) -  \eta(p) \xv) + \alpha(p)\Cm [\uv(p) - \xv] \non
  &=&  -w(p) \Cm(\xv-\fv(p))   \,. 
\end{eqnarray}
In the above equation, we used the expressions of $\gammav(p)$ and
$\Gammam(p)$ in~\eqref{eq:J1} and~\eqref{eq:Gammap}, respectively, and
defined $\gammav(p) = \Cm \betav(p)$ and $\Gammam = \eta(p)\Cm$
where $\eta(p) = \int_q\zeta(p,q)\rho_0(q)\dd q$ and $\betav(p) = \int_\yv\int_q\zeta(p,q)\yv \rho(q,\yv)\dd^N\yv \dd q$.
Moreover, we defined
$w(p) = \bar{\alpha}(p)\eta(p)+\alpha(p)$ and
$\fv(p) = \frac{\bar{\alpha}(p)\betav(p)+\alpha(p)\uv(p)}{w(p)}$.

Interestingly, if vector $\betav(p)$ is given, then $\muv(p,\xv)$ does not
depend on $\rho(p,\xv)$ any longer, rather it depends only on
the personality $p$, the opinion vector $\xv$, and the initial
distribution $\rho_0(p)$.  Furthermore, observe that $\phiv(p)=\frac{\beta(p)}{\eta(p)}$ can
be expressed as the fixed point of the multidimensional Fredholm
equation:
\begin{eqnarray} \phiv(p) &=&
  \underbrace{\int_q\frac{\zeta(p,q)\alpha(q)\uv(q) \rho_0(q)
    }{\eta(p) w(q)}\dd q}_{\hv(q)} \nonumber \\ 
&& \quad + \int_q\underbrace{\frac{\zeta(p,q)\bar{\alpha}(q)\eta(q)}{\eta(p)
      w(q)}}_{\Phi(p,q)}\phiv(q) \rho_0(q)\dd q \nonumber \\ 
&=& \hv(p) + \int_q\Phi(p,q)\phiv(q) \rho_0(q)\dd q \label{eq:beta_Fredholm-a}
\end{eqnarray}
(See Appendix~\ref{app:C} in the Supplemental Material for
details.)

From~\eqref{eq:mu_steady_state}, we observe that if $-w(p)\Cm$ is
stable for every $p$, the stationary solution of the FP equation under
steady-state conditions is given by~\cite{risken1996fokker}:
\begin{equation}\label{eq:rho_W}
 \rho(p,\xv) = \frac{\rho_0(p)}{\sqrt{2 \pi |\Wm(p)|}} \ee^{-1/2(\xv-\fv(p))\Tran \Wm(p)^{-1}(\xv-\fv(p))}
\end{equation}
where $\Wm(p)$ is the unique solution of the Lyapunov equation $\Am(p)\Wm(p)+\Wm(p)\Am(p)\Tran=-\Dm$ and can be expressed as:
\[
 \Wm(p)=-\int_0^\infty \ee^{\Am(p)\tau}\Dm \ee^{\Am(p)\Tran\tau}\dd
 \tau \,.
\]
A more explicit expression of $\rho(p,\xv)$ can be obtained when $\Cm$ is symmetric.
Let us write $\muv(p,\xv)$ as 
\begin{equation} \label{eq:potential}
  \muv(p,\xv) = -\nabla_{\xv} \left[\underbrace{\frac{w(p)}{2}(\xv-\fv(p))\Tran\Cm(\xv-\fv(p))}_{V(p,\xv)}\right]
\end{equation}
where $V(p,\xv)$ is a potential.  If $\Dm$ is symmetric, it can be
written as $\Dm = \Qm\Sigmam\Qm\Tran$ where $\Qm$ is an orthogonal
matrix and $\Sigmam$ is diagonal.  The multi-dimensional FP equation
in~\eqref{eq:fokker-planck-multi} can be rewritten as
 \begin{eqnarray} \label{eq:fokker-planck-multi2}
-\nabla_{\xv}\Tran \left[\nabla_{\xv}V(p,\xv) \rho(p,\xv)\right] = \frac{1}{2}\Trace\left\{\Dm \Hm_{\xv}(\rho(p,\xv))\right\}
\end{eqnarray}
where $\Hm_{\xv}(\rho(p,\xv))$ is the Hessian matrix of $\rho(p,\xv)$
 with respect to the variable $\xv$.  By defining
$\Lm = \Qm\Sigmam^{-1/2}$, we then have $\Lm\Tran\Dm\Lm=\Id$.  Now
consider a new system of coordinates, $\yv$, such that $\xv =
\Lm\yv$. Then, for any twice differentiable function $f(\xv)$, we have
$\nabla_{\xv}f(\xv) = \Lm \nabla_{\yv}f(\Lm\yv)=\Lm
\nabla_{\yv}f_1(\yv)$ and
$\Hm_{\xv}(f(\xv)) = \Lm \Hm_{\yv}(f(\Lm\yv))\Lm\Tran = \Lm
\Hm_{\yv}(f_1(\yv))\Lm\Tran$. By replacing these expressions
in~\eqref{eq:fokker-planck-multi2}, we obtain
 \begin{eqnarray} %\label{eq:fokker-planck-multi3}
\minus\nabla_{\yv}\Tran\Lm\Tran \left[\Lm \nabla_{\yv}V_1(p\mathord{,}\yv) \rho_1(p\mathord{,}\yv)\right] = \frac{\Trace\left\{\Dm \Lm \Hm_{\yv}(\rho_1(p\mathord{,}\yv))\Lm\Tran\right\}}{2}\nonumber
\end{eqnarray}
which, after some algebra, reduces to
\begin{eqnarray} \label{eq:fokker-planck-multi32}
-\nabla_{\yv}\Tran \left[\Sigmam^{-1}\nabla_{\yv}V_1(p,\yv)\rho_1(p,\yv)\right] 
 &=&  \frac{1}{2}\Trace\left\{\Hm_{\yv}(\rho_1(p,\yv))\right\} \non
 &=& \frac{1}{2}\nabla^2_{\yv}\rho_1(p,\yv) \,.
\end{eqnarray}

In the above equation, $V_1(p,\yv)$ has the same structure as
$V(p,\xv)$ in~\eqref{eq:potential} and,
thus,~\eqref{eq:fokker-planck-multi32} is a FP equation in standard
form whose solution is given by the following
%Gibbs  
 distribution:
\begin{equation}\label{gibbs} 
  \rho_1(p,\yv)=\frac{\ee^{-2 V_1(p, \yv)}}{\int\int\ee^{-2 V_1(q, \zv)} \mathrm{d}q \mathrm{d} \zv}.
\end{equation}
%with $K$ being the normalization constant: $K=\int\ee^{-2 V_1(p, \yv)} \mathrm{ d} \bf{y}$. 
% {\color{red} PERCHE' E' in
%  ROSSO?  When $\zeta(p,p')$ can assume negative values, system
%  stability, is not guaranteed in general.  From~\eqref{gibbs} it is
 % of immediate verification that
%  $\lim_{\|\yv \| \to \infty} V_1(p, \yv)=\infty$ for every $p$
%  constitutes a necessary and sufficient condition for stability. }
Therefore the steady state solution of the FP equation can be
expressed as a Gibbs distribution~\eqref{gibbs} associated with a
potential $V_1(p,\yv)$.
   
{\bf Remark 1:} Note that the same expression can be obtained from the
transient solution for $t\rightarrow \infty$ under certain
conditions. Specifically, consider~\eqref{eq:fixed_point}-\eqref{eq:gamma1} and their Laplace
transforms~\eqref{eq:fixed_point_transform}-\eqref{eq:fixed_point_transform_1}; 
using the final-value theorem, we get
\begin{equation}
\lim_{t\rightarrow \infty} \gammav(p,t)  = \lim_{s \to 0} s
\widehat{\gammav}(p,s) \,.
\end{equation}
Assuming that both $\Psim$ and $\Xim(p)$ are Hurwitz-stable for all $p$, then
$\lim_{t\rightarrow \infty} \gammav_0(p,t)  = \lim_{s \to 0} s \widehat{\gammav}_0(p,s) = 0$ and
\begin{align}
&& \lim_{t\rightarrow \infty} \gammav_1(p,t) = \lim_{s \to 0} s \widehat{\gammav}_1(p,s) \nonumber \\
&=& \int_q  \Zm(p,q) \rho_0(q)  \alpha(q)\Xim^{-1}(q) \Cm \uv(q)  \dd q\,. 
\end{align}
So doing, the stationary solution satisfies the integral equation
\begin{eqnarray}\label{eq:gammav_infinity} 
\lim_{t\rightarrow \infty} \gammav(p,t) &=& \lim_{t\rightarrow \infty} \gammav_1(p,t)  \nonumber \\
&&\hspace{-10ex} +\int_q \Zm(p,q)  \rho_0(q) \bar{\alpha}(q)  \Xim^{-1}(q) \lim_{t\rightarrow \infty} \gammav(p,t) \dd q\,. 
\end{eqnarray}
Note that the same observations made for
\eqref{eq:fixed_point_transform} hold also for the stationary
solution~\eqref{eq:gammav_infinity}.

{\bf Remark 2:} When $t\to \infty$, the expression of the
average in \eqref{eq:mv_x} becomes:
\begin{eqnarray}
  &&\hspace{-6ex}\mv(p,\cdot,\infty)\non
  &=&\hspace{-2ex} \lim_{t\to \infty} \ee^{-\Xim(p) t} \yv + \alpha(p)\Xim^{-1}(p) \left(\Id_N\minus\ee^{-\Xim(p) t} \right) \Cm \uv(p)\non
      && \qquad+ \bar{\alpha}(p) \int_0^{t} \ee^{-\Xim(p) (t-\tau)} \gammav(p,\tau) \dd\tau \non
  &=&\hspace{-2ex} \alpha(p)\Xim^{-1}(p)\Cm \uv(p)\non
      && \qquad+\lim_{t\to \infty}  \bar{\alpha}(p) \int_0^{\infty} u(t-\tau)\ee^{-\Xim(p) (t-\tau)} \gammav(p,\tau) \dd\tau \non
  &\stackrel{(a)}{=}&\hspace{-2ex} \alpha(p)\Xim^{-1}(p)\Cm \uv(p)\non
     && \qquad +\bar{\alpha}(p)\lim_{s\to 0} s \Lc\left\{u(t)\ee^{-\Xim(p) t}\right\} \Lc\left\{\gammav(p,t)\right\} \non
  &=&\hspace{-2ex} \alpha(p)\Xim^{-1}(p)\Cm \uv(p)+  \bar{\alpha}(p)\Xim^{-1}(p) \lim_{s\to 0} s\widehat{\gammav}(p,s) \non
  &=&\hspace{-2ex} \Xim^{-1}(p)\left[\alpha(p)\Cm\uv(p)+\bar{\alpha}(p)\gammav(p,\infty)\right] \label{eq:m_infinity}
\end{eqnarray}
where in $(a)$ we applied the final value theorem and we noted that
the integral can be written as the convolution of two functions,
thus its Laplace transform is the product of the transforms of the aforementioned 
functions.

{\bf Remark 3:} When $\rho_0(p)$ is discrete, letting
$t\to \infty$ in \eqref{eq:gamma_t}, we obtain an expression of the
steady state distribution as: $\underline{\gammav}(\infty)= \underline{\Zm}  \Psim^{-1} \Pm_1 \Cm \underline{\uv}$.

At last, observe that the following result holds:
\begin{theorem}  \label{theo2}
	i) The Fredholm equation 
	defining $\phi(p)$ (given explicitly in \eqref{eq:beta_Fredholm-a})  admits a unique
	solution which is Lipschitz-continuous.  ii) The solution $\phi(p)$ of the Fredholm
	equation \eqref{eq:beta_Fredholm-a}, under any distribution
	$\rho_0(p)$, which is continuous at every point in $p$, is the
	uniform limit of solutions $\phi_n(p)$, obtained by replacing
	distribution $\rho_0(p)$ with its discrete approximation $\rho_n(p)$
	whose mesh-size is $\frac{1}{n}$.
\end{theorem}
The proof is provided in Appendix \ref{app:C} in the Supplemental Material.

\section{Summary} \label{sec:summary}
For the sake of clarity, here we summarize the main steps and analytical tools that we used in our derivations.
\begin{itemize}
   \item We first adopted the mean-field approach to model the opinion dynamic evolution through
     a continuous distribution function whose expression can be obtained by solving a FP equation;  
   \item Then, by taking the Fourier transform of the FP equation and using the  method of characteristics,
     we rewrote it as a system of first-order partial-differential
     equations;
   \item Such a system was solved and the final solution
     was obtained by the Fourier inverse transform;
   \item The conditions ensuring the system stability were
     derived for the matrices characterizing the system, by exploiting the
     definition of Hurwitz stability;
   \item Finally, we carried out the steady state analysis starting from the FP
     equation and letting $t\to \infty$. We obtained an expression for
     $\muv(p,\xv)$ which is a function of quantities that can be
     computed by solving a multidimensional Fredholm equation. 
   \end{itemize}

\section{Numerical results\label{sec:results}}

In this section, we show some numerical examples, which shed light on the impact of the model parameters on the stationary state of opinions as well as on their dynamics. In the following, we will always consider a uniform distribution of personalities in the range $[-1,1]$ and a two-dimensional opinion space, i.e., $\xv = [x_1,x_2]^{\Tran}$. 

\subsection{Sensitivity of the stationary distribution on the model parameters}

In this first subsection, we evaluate the impact of the noise variance, the prejudice, and the coupling matrix $\Cm$, on the stationary distribution.

\begin{figure}[tb]
  \centering
\includegraphics[width=1.0\columnwidth]{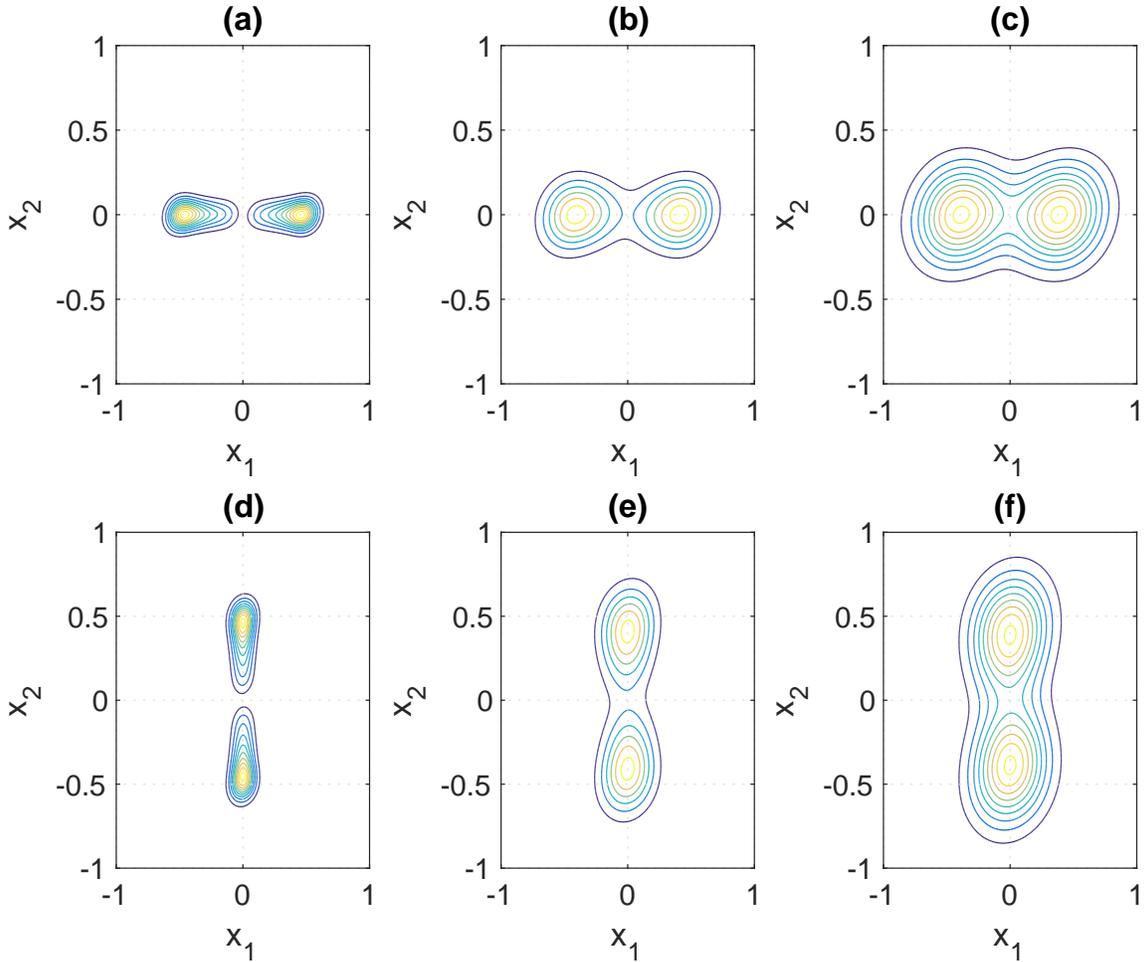}
\caption{Contour lines of stationary opinion distribution for
 $\alpha = 0.01$ and $\zeta(p,q)$ as in \eqref{eq:zeta_proximity}. Plots (a), (b), (c): prejudice as in \eqref{eq:prejudice_1}. Plots (d), (e), (f): prejudice as in \eqref{eq:prejudice_2}. Plots (a), (d):   $\sigma_n^2 = 10^{-3}$. Plots (b), (e): $\sigma_n^2 = 5 \times 10^{-3}$; plots (c), (f): $\sigma_n^2 = 10^{-2}$. \label{fig:Noise_effect}}%end caption
\vspace{-3mm}
\end{figure}

We first show the effect of noise variance $\sigma^2_n$ and the prejudice $\uv$. To this purpose, we consider an asymmetric coupling matrix 
\beq \label{eq:corr_mat_asym}
\Cm = \left[
\begin{array}{cc}
1 & \rho \\
\epsilon & 1
\end{array}
\right]
\eeq 
where $\rho = 0.3$ and $\epsilon$ is a very small positive number
(namely, $10^{-10}$ while obtaining the results)\footnote{We have used
  $\epsilon$ as an approximation to $0$, since setting $\epsilon = 0$
  would yield a non-diagonalizable matrix $\Cm$.}. Such
  model is well suited for a case where the reciprocal influence of the
  opinions on two subjects is unidirectional (e.g., from subject 2 to subject
  1): a possible example could be the appreciation of the government
  action (subject 1), and the opinion on the right level of taxation
  (subject 2), where subject 2 is more likely to affect subject 1 than
  vice versa.  We assume a constant level of stubbornness $\alpha(p)
= 0.01$, while the strength of opinion interaction is given by
\beq \label{eq:zeta_proximity}
\zeta(p,q) = \frac{1}{1+|p-q|^2}
\eeq
 i.e., interactions are stronger between agents with similar personalities, a model which we will refer to as \emph{proximity} model.
Fig.\,\ref{fig:Noise_effect} shows the contour lines of the
stationary opinion distribution for different values of $\sigma_n^2$,
for two different prejudice scenarios. In the top row, the prejudice is given by
\beq \label{eq:prejudice_1}
\uv(p) = \left\{
\begin{array}{cc}
[-1, 0]\Tran, & p<0 \\
{[1, 0]\Tran},  & p \geq 0
\end{array}
\right.
\eeq 
while in the bottom it satisfies
\beq \label{eq:prejudice_2}
\uv(p) = \left\{
\begin{array}{cc}
[0,-1]\Tran, & p<0 \\
{[0,1]\Tran},  & p \geq 0 \,.
\end{array}
\right.
\eeq
In both rows, we set $\sigma_n^2 = 10^{-3}$ for the left plot,
$\sigma_n^2 = 5 \times 10^{-3}$ for the center plot, and $\sigma_n^2 =
10^{-2}$ for the right plot. Observe that  the stationary distribution features two peaks, corresponding to the two different prejudice points, with the same height and width. The width increases with increasing noise variance, for the highest noise variance, the peaks start to merge. 
%It is worth noting that, while in the top-row plots the peaks are aligned on the $x_1$ axis, in those of the bottom row, the dynamic on the $x_2$ axis is affected by that on the $x_1$ axis, and the two peaks are tilted with respect to the $x_2$ axis. 

\begin{figure}[tb]
  \centering
\includegraphics[width=1.0\columnwidth]{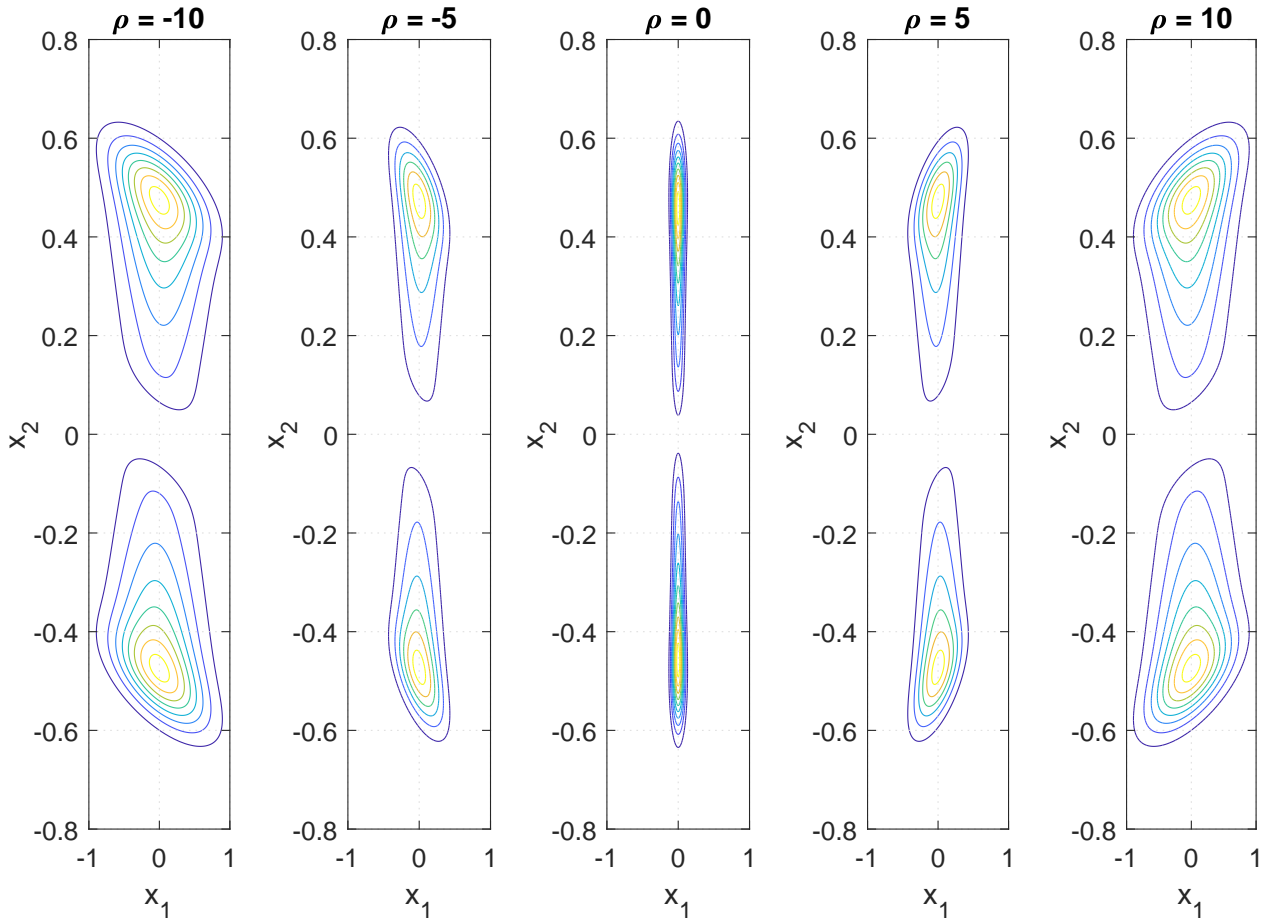}
\caption{Contour lines of stationary opinion distribution for
 $\alpha = 0.01$, $\sigma_n^2 = 10^{-3}$,  and $\zeta(p,q)$ as in  \eqref{eq:zeta_proximity}. Coupling matrix given by \eqref{eq:corr_mat_asym}. From left to right: $\rho = -10, -5, 0 , 5, 10$. \label{fig:Corr_effect}}%end caption
\vspace{-3mm}
\end{figure}

Next, in Fig.\,\ref{fig:Corr_effect} we investigate the effect of topic correlation, as expressed by the coupling matrix $\Cm$. We use the same interaction strength as per \eqref{eq:zeta_proximity}, and the prejudice given in \eqref{eq:prejudice_2}. Also, as before, $\alpha(p) = 0.01$ and $\sigma_n^2 = 10^{-3}$. Finally, we consider the coupling matrix as in \eqref{eq:corr_mat_asym}, with $\rho \in \{-10,-5,0,5,10\}$. 
For $\rho=0$ there is no interaction between the two opinion
components; changing $\rho$ does not have any effect on the mean of the stationary distribution for each personality (\eqref{eq:mv_x}), leading to peaks whose locations are essentially invariant with respect to $\rho$. Notice that the distribution of $x_2$ is independent of $\rho$. Moreover, due to the symmetric scenario, in all cases the stationary distribution shows a reflectional symmetry around the origin. Finally, changing $\rho$ into $-\rho$ has, in the considered scenario, the effect of reflecting the stationary distribution around the $x_2$-axis, or, in other words, of changing $x_1$ into $-x_1$. From the system point of view, the effect of different correlation values implies a larger share of agents that have strong positive opinions on both subjects (positive correlation value) or one strong positive and one strong negative opinion (negative correlation value). Going back to the previous practical example, agents which are in favor of a low level of taxation will judge more positively or more negatively the government action, depending on the correlation coefficient sign.
%This is true because the dynamics of opinion $x_2$ is not affected by such a change, being both prejudice and interactions symmetric with respect to the origin, while for the dynamics of $x_1$ the effect of the sign changes in $\rho$ and $x_2$ compensate each other.      

\subsection{Community-based scenario, the influence of \protect$\Zm$ and the insurgence of instability}

We now assess the impact of the interaction strength matrix $\Zm$ on
the stability of opinions.  We consider a different scenario, in which
there are $M$ personalities ($M$ even), all with the same stubbornness
level $\alpha = 0.01$, organized in two communities and with an
interaction similar to that considered in Example 1, but with $N=2$,
i.e., \beq \label{eq:zeta_community} \underline{\Zm} = \left(\left[
\begin{array}{cc}
\zeta_1 & - \zeta_2 \\
- \zeta_2  & \zeta_1
\end{array} \right] \otimes \mathbf{1}_{M/2} \right) \otimes \Cm
\eeq
with $\zeta_1 = 1$ and $\zeta_2 $ used as a parameter. We consider
$\Cm$ as given by \eqref{eq:corr_mat_asym} with $\rho = 1$, the
prejudice as given in \eqref{eq:prejudice_2}, and $\sigma_n^2
=10^{-3}$.
Also, $\rho_0(p)$ is given by~\eqref{eq:rho0} with $r_i=1/M$ for all $i$. 
As a possible practical example of such a scenario, we can
think of two religious sects, say Bogumils and Cathars, which have
generally different views on two topics,  represented by the two
opinions $x_1$ and $x_2$. 

It is not difficult to show that the stability region boundaries are
the same as for Example 1, since $\Cm$ is Hurwitz-stable.
Furthermore, we can derive the asymptotic expression of the mean in~\eqref{eq:m_infinity}
by exploiting Remark 3 in Section~\ref{sec:steady_state} and writing from~\eqref{eq:Gammap} and~\eqref{eq:Xi}
$\underline{\Xim}=\beta\Id_M\otimes \Cm$, where $\beta=\bar{\alpha}\frac{1-\zeta_2}{2}+\alpha$.
If $\beta>0$, through some algebra and by applying the properties of the Kronecker product, we obtain 
\beq \label{eq:asym_mean}
\mv(p,\cdot,\infty) = \frac{\alpha}{\alpha - \overline{\alpha}
  \zeta_2} \uv(p). \eeq
Thus, if $\zeta_2 < \alpha/\overline{\alpha} \simeq 0.0101$, the system is
stable.

\begin{figure}[tb]
  \centering
\includegraphics[width=1.0\columnwidth]{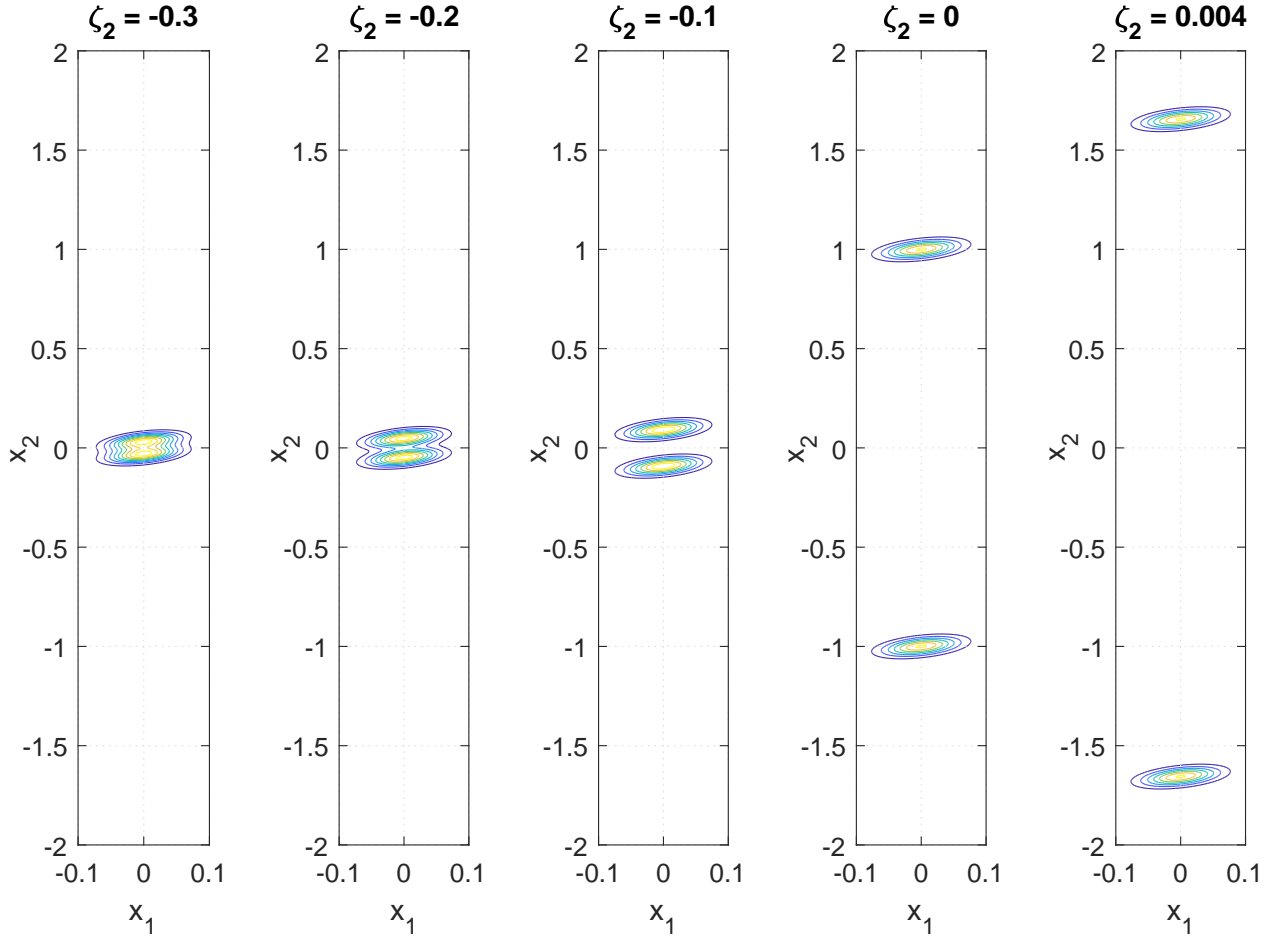}
\caption{Stationary opinion distribution for
 $\alpha = 0.01$, $\sigma_n^2 = 10^{-3}$, $\Zm$ as in  \eqref{eq:zeta_community} ($\zeta_1 = 1$), for several values of $\zeta_2$ in the stability region. Coupling matrix given by \eqref{eq:corr_mat_asym} with $\rho = 1$. From left to right: $\zeta_2 = -0.3,-0.2,-0.1,0,0.004$. \label{fig:Comm_stable}}%end caption
\vspace{-3mm}
\end{figure}

Fig. \ref{fig:Comm_stable} shows the stationary distribution
for increasing value of $\zeta_2$. As it can be seen, Bogumils and
Cathars merge because of the attractive forces for $\zeta_2 = -0.2$ or
lower, while, for $-0.1 < \zeta_2 < 0$, notwithstanding their
reciprocal attraction, they remain separated because of the effect of
the prejudice.  For $ 0 < \zeta_2 < \alpha/\overline{\alpha}$, the two
communities repel each other, but this repulsion is not strong enough
to win the effect of prejudice, so stability is preserved while the
means grow larger and larger for
$\zeta_2 \uparrow \alpha/\overline{\alpha}$.

\begin{figure}[tb]
  \centering
\includegraphics[width=1.0\columnwidth]{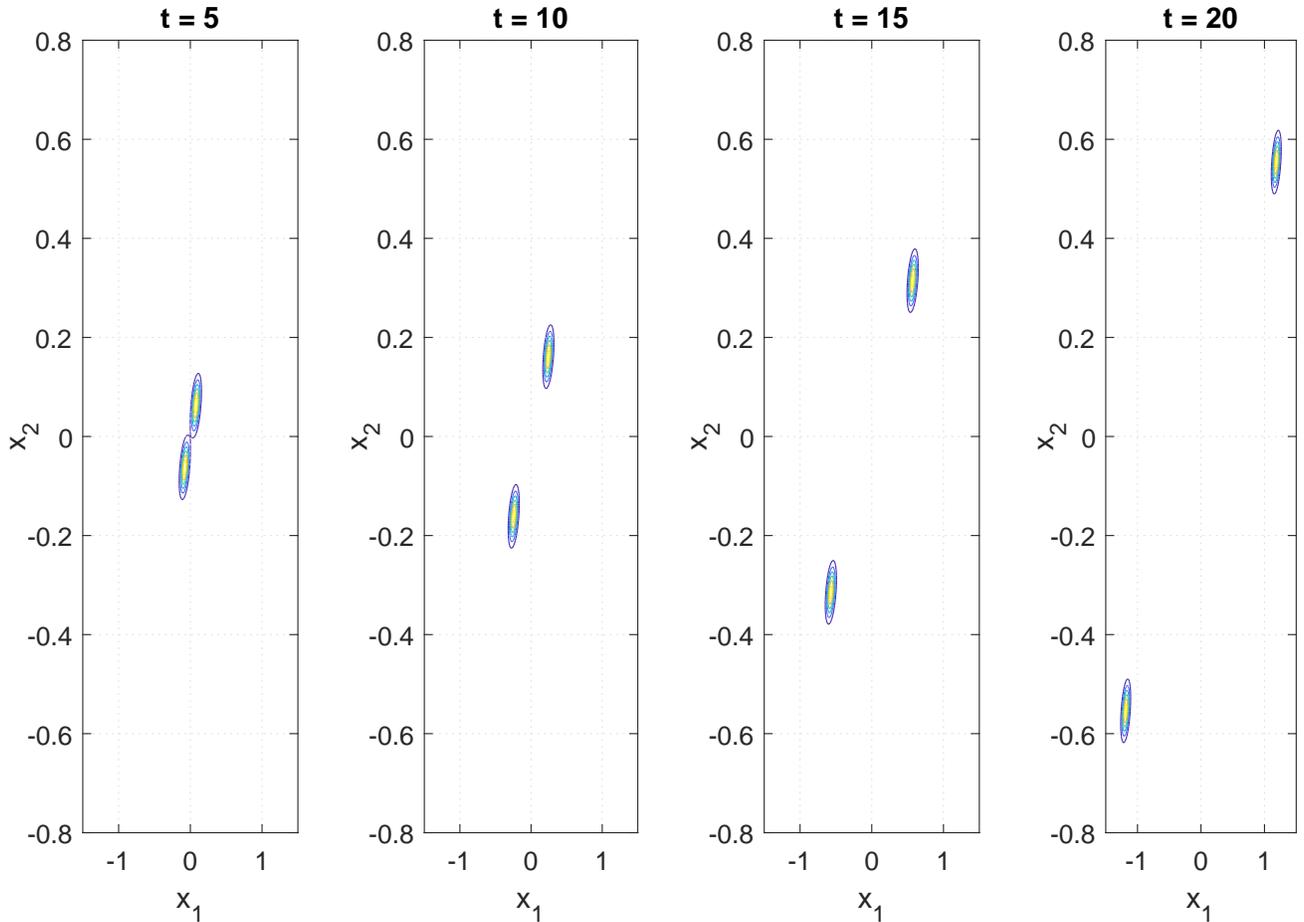}
\caption{Contour lines of opinion distribution for
 $\alpha = 0.01$, $\sigma_n^2 = 10^{-3}$, $\Zm$ as in  \eqref{eq:zeta_community} ($\zeta_1 = 1$, $\zeta_2 = 0.1$). Coupling matrix given by \eqref{eq:corr_mat_asym} with $\rho = 1$. From left to right: $t = 5,10,15,20$. \label{fig:Comm_unstable}}%end caption
\vspace{-3mm}
\end{figure}

For $\zeta_2 > \alpha/\overline{\alpha}$, the system is not stable
anymore, and \eqref{eq:asym_mean} does not hold. In particular, when
$\zeta_2 < \zeta_1 + 2 \alpha/\overline{\alpha} = 1.0202$,   the
intra-community attraction and the inter-community repulsion have such
a relative strength that the system experiences type-II instability:
the communities are preserved but their respective means tend to
diverge. In our example, the two religious sects are so enemy of each
other, to radicalize their views while retaining a strong identity
within themselves, giving rise to religious fanaticism. 
We now look at the time evolution of opinions in the case where, for all personalities, opinions start deterministically from the origin. Fig.\,\ref{fig:Comm_unstable} shows the opinions at time instants $t = 5,10,15,20$, for $\zeta_2 = 0.1$. Notice that, because of the correlation, the two communities diverge along the bisector of the I-III quadrants.

Finally, for $\zeta_2 > \zeta_1 + 2 \alpha/\overline{\alpha}$, the
inter-community repulsion prevails on intra-community attraction, the
system experiences type-I instability, and the variance within the two
communities also grows to infinity. In other words, Bogumils and
Cathars dissolve themselves into heterogeneous crowds not having
definite views on the two topics of
interest. Fig.\,\ref{fig:Comm_unstable} shows the opinions at time
instants $t = 0.15,0.3,0.45,0.5$ for $\zeta_2 = 10$, again when the
opinions all start deterministically from the origin. The dynamic is
similar, but faster than in the previous case, and the communities
expand, so that, for $t$ sufficiently large, their boundaries disappear.    

 \begin{figure}[tb]
  \centering
\includegraphics[width=1.0\columnwidth]{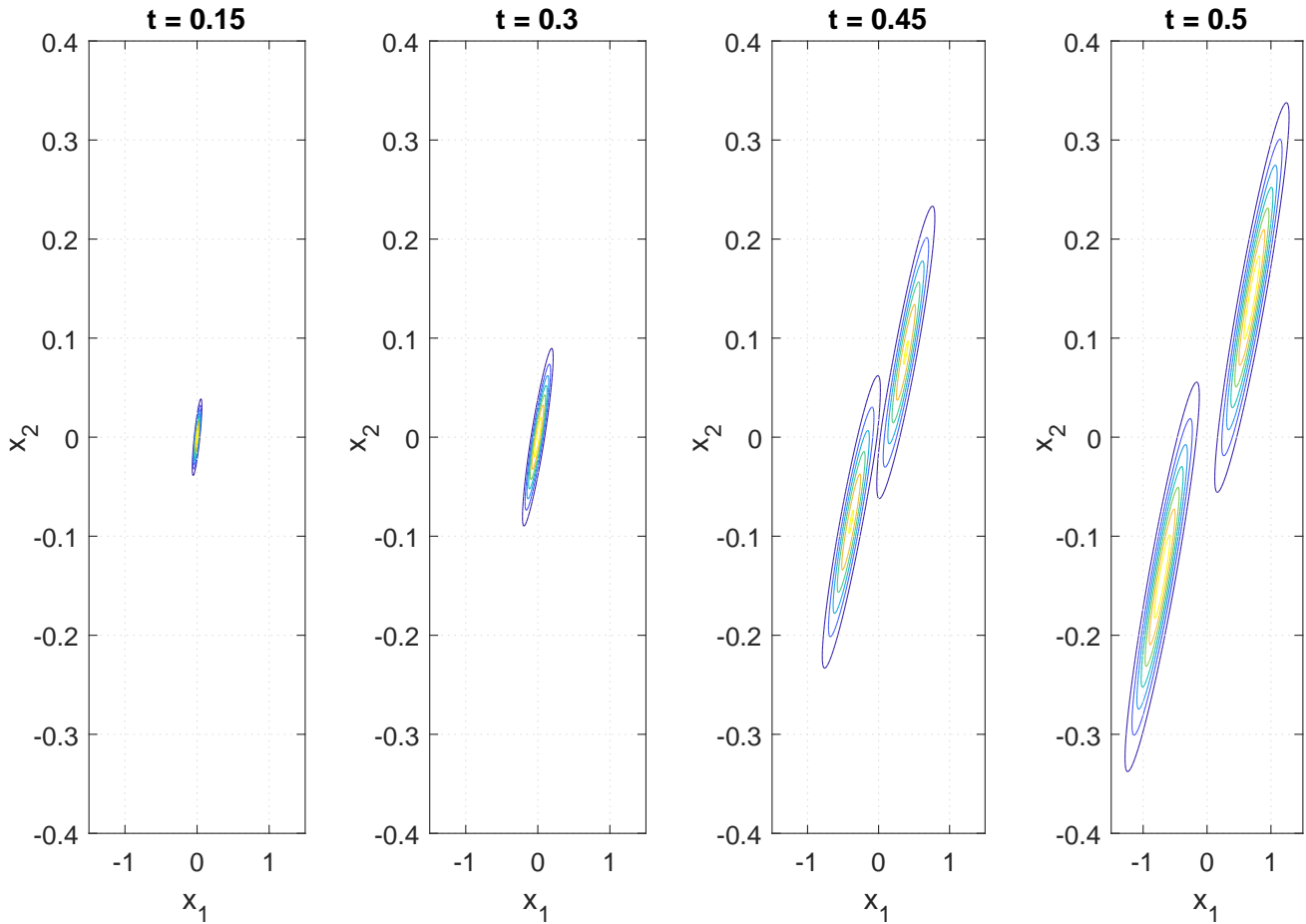}
\caption{Contour lines of opinion distribution for
 $\alpha = 0.01$, $\sigma_n^2 = 10^{-3}$, $\Zm$ as in  \eqref{eq:zeta_community} ($\zeta_1 = 1$, $\zeta_2 = 10$). Coupling matrix given by \eqref{eq:corr_mat_asym} with $\rho = 1$. From left to right: $t = 0.15,0.3,0.45,0.5$. \label{fig:Comm_unstable_2}}%end caption
\vspace{-3mm}
\end{figure}

\subsubsection{Finite-network behavior}

\begin{figure}[tb]
  \centering
\includegraphics[width=1.0\columnwidth]{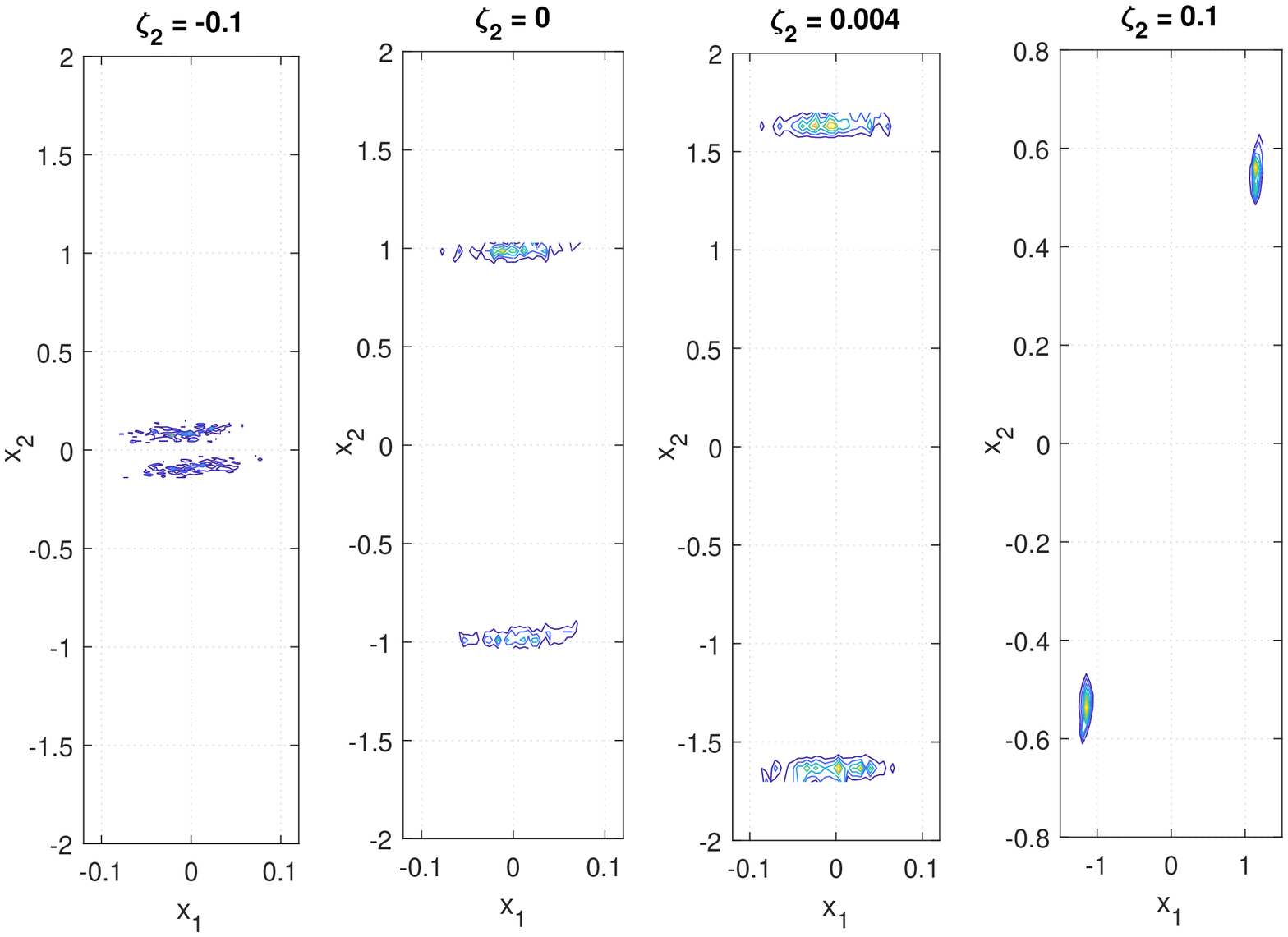}
\caption{Contour lines of simulated opinion distribution for
 $U = 500$, $\alpha = 0.01$, $\sigma_n^2 = 10^{-3}$, $\Zm$ as in  \eqref{eq:zeta_community} ($\zeta_1 = 1$), where the first half of users belong to community 1, while the second half to community 2. Coupling matrix given by \eqref{eq:corr_mat_asym} with $\rho = 1$. From left to right: $\zeta_2 = -0.1, 0, 0.004, 0.1$. For the rightmost graph, $t=20$. \label{fig:Comm_simul}}%end caption
\vspace{-3mm}
\end{figure}

In order to assess the validity of the mean-field approach, in
Fig. \ref{fig:Comm_simul} we show the opinion distribution behavior
for a finite set of $U = 500$ agents, by solving numerically
\eqref{eq:x_i}. In particular, we consider the same model as in
Figs. \ref{fig:Comm_stable}-\ref{fig:Comm_unstable} for
$\zeta_2 \in \{-0.1, 0, 0.004, 0.1\}$. The first half of the agents
are Bogumils, while the others are Cathars. For the first three values
of $\zeta_2$, which correspond to a stable system, we show the
stationary opinion distribution.  For $\zeta_2 = 0.1$, we show the
distribution at time $t = 20$. As it can be seen, the results in 
Figs. \ref{fig:Comm_stable}-\ref{fig:Comm_unstable} match those in Fig.
\ref{fig:Comm_simul}, indicating that the asymptotic analysis
well represents the opinion dynamics of relatively small populations. 

\subsection{Rotational effects}

\begin{figure*}[tb]
  \centering
\includegraphics[width=1.0\textwidth]{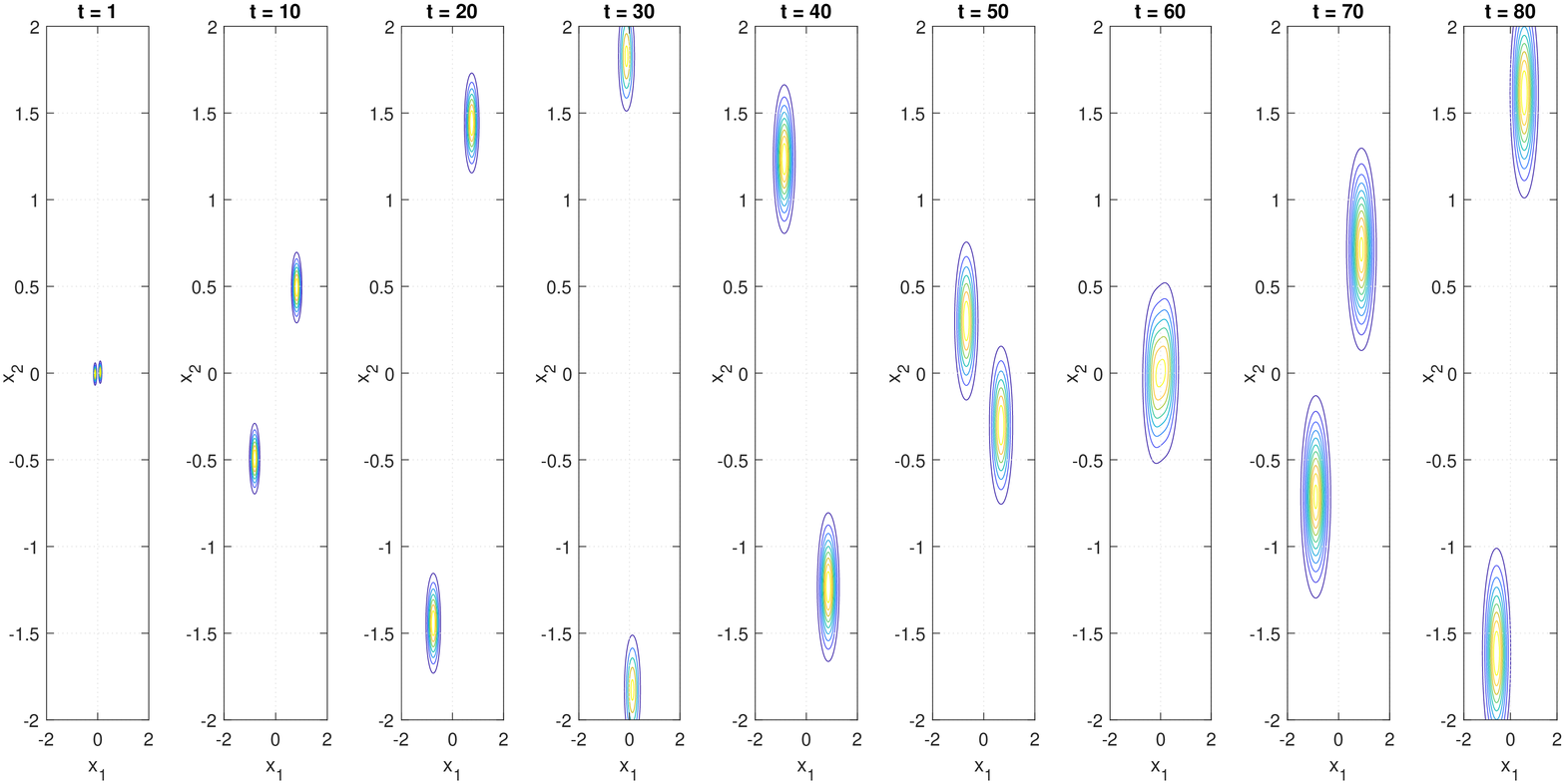}
\caption{Contour lines of opinion distribution for
 $\alpha = 0.01$, $\sigma_n^2 = 10^{-3}$, $\Zm$ as in  \eqref{eq:zeta_community} ($\zeta_1 = 1$, $\zeta_2 = -0.1$). Coupling matrix given by \eqref{eq:corr_mat_complex}. From left to right: $t = 1,10,20,30,40,50,60,70,80$. \label{fig:Comm_rotating}}%end caption
\vspace{-3mm}
\end{figure*}

We now consider the case in which the dynamics is stable in
both dimensions, but the effect of the coupling matrix yields
instability (of Type I). Consider again the two-community scenario
given by \eqref{eq:zeta_community}, but with $\zeta_1 = 1$ and
$\zeta_2 = -0.1$. Note that, in the scalar case, this would be a stable scenario. However, let us now consider a coupling matrix given by
\beq \label{eq:corr_mat_complex}
\Cm = \left[
\begin{array}{cc}
0 & 1 \\
-1 & 0
\end{array}
\right]
\eeq
which is easily seen to have eigenvalues equal to $\pm j$. This is
quite an artificial scenario, since this coupling matrix has rather an
ad-hoc shape, to which it is difficult to associate any real-world situation. All the other parameters are the same as in the previous example, except for the prejudice, which is given by
\beq \label{eq:prejudice_22}
\uv(p) = \left\{
\begin{array}{cc}
[0,-10], & p<0 \\
{[0,10]},  & p \geq 0 \,.
\end{array}
\right.
\eeq
 It is easy to see that, for the $i$-th discrete personality, the covariance matrix in \eqref{eq:covariance} is given by:
\beq
\Sigmam \left( p_i,t \right) = \sigma^2_n t \Id_2
\eeq   
independently from $i$. Note that such covariance matrix does not
reach any finite limit for $t \rightarrow \infty$, hence the system is
unstable. Fig.\,\ref{fig:Comm_rotating} shows the temporal evolution of the opinion distribution, for $t = 1,10,20,30,40,50,60,70,80$. The peaks corresponding to the two communities widen along time, as expected, while their means undergo a rotation around the origin, at some instants (such as at $t = 60$) making the peaks temporarily merge.

\section{Conclusions\label{sec:conclusions}}
Several analytical models representing the dynamics of social opinions
have been proposed in the literature. By drawing on this body of work,
we developed a model that accounts for the individual endogenous
process of opinion evolution as well as for the possible adversarial
behavior of individuals. Importantly,  our model also represents the interdependence
between  opinions on different, yet correlated, subjects.  Under
asymptotic conditions on the size of the individuals' population, we obtained the time
evolution of the opinions distribution as the solution of a
multidimensional Fokker-Planck equation. We then discussed the
stability conditions and derived the steady-state solution.  Our
numerical results match the stability conditions we obtained and show
interesting phenomena in collective belief dynamics.

\bibliographystyle{IEEEtran}
\bibliography{Refs_MD}

\newpage
%\onecolumn
\appendices

\section{Derivation of~\eqref{eq:solution}\label{app:solution}}
By using~\eqref{eq:mu_x2} and by neglecting dependencies on the system
variables unless strictly necessary, we can rewrite the FP equation
in~\eqref{eq:fokker-planck} as:
\begin{eqnarray} 
  \frac{\partial }{\partial t} \rho(\xv)
  &=& \sum_{n=1}^N \frac{\partial }{\partial x_n}  \left[(\ev_n\Tran \Xim \xv - \phi_n \rho(\xv)\right] + 
      \frac{1}{2} \sum_{m,n=1}^N D_{mn} \frac{\partial^2 }{\partial x_m \partial x_n}  \rho(\xv)\label{eq:start}
\end{eqnarray}
where $\ev_n$ is the vector with all zero components but the $n$-th one equal to 1.
Then, by taking the Fourier transform of~\eqref{eq:start} with respect to $\xv$,
we obtain 
\begin{equation} 
\frac{\partial }{\partial t} \widehat{\rho}(\nuv) + \nuv\Tran \Xim \nabla_{\nuv} \widehat{\rho}(\nuv) =
-\left[j 2\pi \nuv\Tran \phiv + 2 \pi^2 \nuv\Tran \Dm \nuv \right]
\widehat{\rho}(\nuv) \label{eq:FPtransformed}
\end{equation}
where $\nuv$ is the variable in the transformed domain, while
$\hat{f}(\nuv)$ is the Fourier transform of the generic function
$f(\xv)$. The operator $\nabla_{\nuv}$ is the gradient with respect to the vector $\nuv$.
Next we transform the above first-order partial-derivative
equation into a system of first-order ordinary differential equations
by using the method of characteristics~\cite{Sarra}.  To this end we
introduce an auxiliary parameter $u$ and consider $t=t(u)$,
$\nuv=\nuv(u)$ and $\widehat{\rho}(p,\nuv(u),t(u)) = \widehat{\rho}(u)$, with initial
conditions $t(0) = 0$, $\nuv(0) = \nuv_0$ and
$\widehat{\rho}(p,\nuv(0),t(0)) =\widehat{\rho}_0(p,\nuv_0) = \widehat{\rho}(0)$.
According to the rule of total derivative, $\frac{\dd \widehat{\rho}}{\dd u}$ can be written as
\[ \frac{\dd \widehat{\rho}}{\dd u} = \frac{\partial \widehat{\rho}
  }{\partial t}\frac{\dd t}{\dd u}+\frac{\dd \nuv\Tran}{\dd
    u}\nabla_{\nuv} \widehat{\rho} \,.\] 

If we set
$\frac{\dd t}{\dd u}=1$ and $\frac{\dd \nuv}{\dd u} = \Xim\Tran\nuv$,
the above total derivative corresponds to the left hand side 
of~\eqref{eq:FPtransformed}. With such setting we
can reduce~\eqref{eq:FPtransformed} to the following system of
differential equations:
\begin{equation}
\left\{
\begin{array}{ccl}
\frac{\dd t}{\dd u} & =&  1 \\
\frac{\dd \nuv}{\dd u} & = & \Xim\Tran \nuv \\
\frac{\dd \widehat{\rho}}{\dd u} & = & -\left[j 2\pi \nuv\Tran \phiv +
                                       2 \pi^2 \nuv\Tran \Dm \nuv
                                       \right] \widehat{\rho} \,.
\end{array}
\right.
\end{equation}
The first two equations are easily solved as $t(u)=u$, and $\nuv(u)  =
\ee^{\Xim\Tran u} \nuv_0$. 
By substituting these solutions in the third equation, the latter can be rearranged as  
\begin{equation}
\frac{\dd \widehat{\rho}(u)}{\widehat{\rho}(u)}  =  -\left[j 2\pi \nuv_0\Tran \ee^{\Xim u} \phiv + 2 \pi^2 \nuv_0\Tran \ee^{\Xim u} \Dm \ee^{\Xim\Tran u} \nuv_0 \right] \dd u
\end{equation}
We now recall that $\phiv = \phiv(t)$ is a function of $t$. Since
$t=u$, by integrating with respect to $u$ the above equation, we get
\begin{equation}
\log \frac{\widehat{\rho}(u)}{\widehat{\rho}(0)}  =  -j 2\pi \nuv_0\Tran  \int_0^{u} \ee^{\Xim w} \phiv(w) \dd w - 2 \pi^2 \nuv_0\Tran \left( \int_0^{u} \ee^{\Xim w} \Dm \ee^{\Xim\Tran w} \dd w \right) \nuv_0 
\end{equation}
The solution for $\hat{\rho}(u)$ is given by
\begin{eqnarray}
\widehat{\rho}(u)  & =& \widehat{\rho}(0) \exp\left\{ -j 2\pi \nuv_0\Tran  \int_0^{u} \ee^{\Xim w} \phiv(w) \dd w - 2 \pi^2 \nuv_0\Tran \left( \int_0^{u} \ee^{\Xim w} \Dm \ee^{\Xim\Tran w} \dd w \right) \nuv_0\right\}
%& =& \delta(p-q) \exp\left\{ -j 2\pi \nuv_0\Tran  \left[\yv +  \int_0^{u} \ee^{\Xim w} \phiv(w) dw\right] - 2 \pi^2 \nuv_0\Tran \left( \int_0^{u} \ee^{\Xim w} \Dm \ee^{\Xim\Tran w} dw \right) \nuv_0\right\}
 \label{eq:rho_u} 
\end{eqnarray}
By substituting $u=t$ and $\nuv_0  = \ee^{-\Xim\Tran t} \nuv(t)$ into \eqref{eq:rho_u}, and after reintroducing all dependencies we finally obtain
\begin{eqnarray}
  \widehat{\rho}(p,\nuv ,t)
  &=& \widehat{\rho}_0(p,\nuv_0)\exp\left\{-j 2\pi \nuv\Tran  \underbrace{\ee^{-\Xim(p) t} \left[\int_0^{t} \ee^{\Xim(p) \tau} \phiv(p,\tau)\dd\tau \right]}_{\widetilde{\mv}(p,t)} - 2 \pi^2 \nuv\Tran \underbrace{\left( \int_0^{t} \ee^{-\Xim(p) \tau} \Dm \ee^{-\Xim(p)\Tran \tau} \dd\tau \right)}_{\Sigmam(p,t)}\nuv\right\} \non
  &=& \widehat{\rho}_0\left(p,\nuv \ee^{-\Xim(p) t}\right)\exp\left\{-j 2\pi \nuv\Tran  \widetilde{\mv}(p,t) - 2 \pi^2 \nuv\Tran \Sigmam(p,t)\nuv\right\} \label{eq:rho_p_nu_t}
\end{eqnarray}
where we used the fact that
\begin{eqnarray}
 \nuv_0\Tran  \left(\int_0^{t} \ee^{\Xim w} \Dm \ee^{\Xim\Tran w} \dd w \right) \nuv_0 
  &=&\nuv \Tran \left(\int_0^{t} \ee^{-\Xim(p) t} \ee^{\Xim w} \Dm \ee^{\Xim\Tran w}\ee^{-\Xim(p) t} \dd w \right) \nuv \non
  &=& \nuv \Tran \left(\int_0^{t} \ee^{\Xim (w-t)} \Dm \ee^{\Xim\Tran (w-t)} \dd w  \right) \nuv \non
  &\stackrel{(a)}{=}& \nuv\Tran\left(  \int_0^{t} \ee^{-\Xim\tau} \Dm \ee^{-\Xim\Tran \tau} \dd \tau  \right) \nuv   
\end{eqnarray}
and in $(a)$ we defined $\tau = t-w$.
Now, by taking the inverse Fourier transform w.r.t. $\nuv$ of~\eqref{eq:rho_p_nu_t}, we get
\begin{equation}
\rho(p,\xv ,t) = \ee^{\Xim(p) t}\rho_0\left(p,\xv\ee^{\Xim(p)t}\right) \star\Gc\left(\xv, \widetilde{\mv}(p,t), \Sigmam(p,t)\right)\label{eq:convolution}
\end{equation}
where the symbol $\star$ represents the convolution operator and
$\Gc\left(\xv, \widetilde{\mv}(p,t), \Sigmam(p,t)\right)$ is the pdf
of the multivariate Gaussian distribution with mean
$\widetilde{\mv}(p,t)$ and covariance $\Sigmam(p,t)$.  Finally, after
a suitable change of variable, and by recalling that $\rho_0(p,\yv) = \rho_0(\yv|p)\rho_0(p)$, we rewrite  the convolution product in~\eqref{eq:convolution} as
\begin{eqnarray}
  \rho(p,\xv,t)
%  &=& \int_{\yv}  \rho_0(p,\yv)\, \Gc\left(\xv, \mv(p,\yv,t), \Sigmam(p,t)\right) \dd^N \yv \non
  &=& \rho_0(p) \int_{\yv}  \rho_0(\yv | p)\, \Gc\left(\xv, \mv(p,\yv,t), \Sigmam(p,t)\right) \dd^N \yv
      \label{eq:convolution2}
\end{eqnarray}
%or, equivalently, as
%\begin{equation}
%  \rho(p,\xv,t) = \int_q \int_{\yv}  \rho_0(q,\yv)\underbrace{\delta(p-q)\,\Gc\left(\xv, \mv(p,\yv,t)%, \Sigmam(p,t)}_{\rho(p,\xv,t|q,\yv)}\right) \dd^N \yv \dd q
%\end{equation}
where
%$\rho(p,\xv,t|q,\yv)$ is the Green function of the system and
%\begin{eqnarray}
$\mv(p,\yv,t) \triangleq  \ee^{-\Xim(p) t}\yv +\widetilde{\mv}(p,t)
= \ee^{-\Xim(p) t} \left[ \yv+\int_0^{t} \ee^{\Xim(p) \tau} \phiv(p,\tau)\dd\tau \right]$. 
%\end{eqnarray}

\section{Proof of Theorem \ref{theo2} \label{app:C}}
\subsection{Preliminaries}
Under steady state conditions, using~\eqref{eq:rho_W} and the definition of $\fv(p)$,
we can rewrite $\betav(p)$ as
\begin{eqnarray}
  \betav(p)
  &=& \int_q\zeta(p,q) \rho_0(q) \int_{\xv} \frac{\xv }{\sqrt{2\pi |\Wm(q)|}}\ee^{-\frac{1}{2}(\xv-\fv(q))\Tran\Wm(q)^{-1}(\xv-\fv(q))}\dd^N\xv \dd q \non
  &=& \int_q\zeta(p,q)  \rho_0(q) \fv(q)\dd q
\end{eqnarray}

By changing variable and considering
$\phiv(p) = \frac{\betav(p)}{\eta(p)}$, we rewrite the above
expression as
\begin{eqnarray} \phiv(p) &=&
  \underbrace{\int_q\frac{\zeta(p,q)\alpha(q)\uv(q) \rho_0(q)
    }{\eta(p) w(q)}\dd q}_{\hv(q)}+
  \int_q\underbrace{\frac{\zeta(p,q)\bar{\alpha}(q)\eta(q)}{\eta(p)
      w(q)}}_{\Phi(p,q)}\phiv(q) \rho_0(q)\dd q \non &=& \hv(p) +
  \int_q\Phi(p,q)\phiv(q) \rho_0(q)\dd q \label{eq:beta_Fredholm}
\end{eqnarray}
which is the Fredholm equation of the second type~\cite{kolmogorov2012functions}.
In the following, we will consider a per-element solution of the above equation.

In order to solve the Fredholm equation, let us denote by $\Ac[\phi](p)$ the following operator
\begin{equation}\Ac[\phi](p) = \int_p \Phi(p,q) \phi(q) \rho_0(q)\dd q\label{eq:A_phi}
\end{equation}
operating on a Banach space of Lipschits continuous functions $\phi(q)$, where $\phi(q)$  
is the generic component of $\phiv(p)$.
We have that $\phi(p)$ is equipped with the following norm: 
\begin{eqnarray}
\| \phi(p) \|_{\rm Lip} &=& c_0 \sup_{\Pc} |\phi(p)|+ c_1 \sup_{p,q \in \Pc, p\neq q} \frac{|\phi(p)-\phi(q)|}{|p-q|} \non 
&=& c_0\|\phi(p)\|_\infty+ c_1 \| \phi(p) \|_{\rm L} 
\end{eqnarray}
with  $c_0,c_1>0$ and $c_0+c_1=1$.
Then we can  rewrite \eqref{eq:beta_Fredholm} as  
\[
 \phi(p)= h(p)+\Ac[\phi](p)
\]
where $h(p)$ is the generic component of $\hv(p)$.
From the above expression, we get
\[
 \phi(p)=(\Ic-\Ac)^{-1}[h](p)
\]
whenever $(\Ic-\Ac)^{-1}[\cdot]$  exists and is continuous over the aforementioned
Banach space.

Now let $R(p)$, $R:\Pc\to [0,1]$ be a continuous initial cumulative distribution
with pdf  $\rho_0(p)$, and 
let $R_n(p)$ be the stepwise approximation of $R(p)$  with meshsize equal to $1/n$. 
Then, using~\eqref{eq:A_phi}, we can write
\[
 \Ac_n[\phi]= \int_\Pc   \Phi(p,q) \phi(q) \dd R_n(q), \quad \Ac[\phi]= \int_\Pc   \Phi(p,q) \phi(q)\dd R(q),
\]
and 
\[
 h_n(p)= \int_{\Pc} \frac{\zeta(p,q) \alpha(q) u(q)}
 {\eta(p) w(q)} \dd R_n(q),  \quad  h(p)= \int_{\Pc} \frac{\zeta(p,q) \alpha(q) u(q)}
 {\eta(p) w(q)} \dd R(q), 
\]
Given the above definitions, we introduce $\phi_n(p)$ as 
\[ \phi_n(p) = \left(\Ic - \Ac_n\right)^{-1}[h_n](p) \,.\] 

\subsection{Main Theorem}
We can now prove  Theorem  \ref{theo2}  whose statement is reported
again below for completeness:
\begin{theorem}
  i) The Fredholm equation \eqref{eq:beta_Fredholm} admits a unique
  solution which is Lipschitz-continuous.  ii) The solution $\phi(p)$ of the Fredholm
  equation \eqref{eq:beta_Fredholm}, under any distribution
  $\rho_0(p)$, which is continuous at every point in $p$, is the
  uniform limit of solutions $\phi_n(p)$, obtained by replacing
  distribution $\rho_0(p)$ with its discrete approximation $\rho_n(p)$
  whose mesh-size is $\frac{1}{n}$.
\end{theorem}

\IEEEproof
  In order to prove the thesis we proceed as follows: 
  \begin{itemize}
  \item i) descends from the properties of operator $\Ac[\cdot]$ over
    the Banach space of Lipschitz-continuous functions equipped with
    norm $\| \cdot \|_{\rm Lip}$, which satisfies
    $\| \Ac[\cdot]\|_{\rm Lip}<1$, since we
    can derive the existence and the continuity of the operator
    $(\Ic-\Ac)^{-1}[\cdot]$ with respect to norm $\| \cdot \|_{\rm
      Lip}$ (see Lemma~\ref{lemma:1} below).
  \item ii) descends from Lemma~\ref{lemma:2} which shows that
    operators $\Ac_n[\cdot]$ converge to the operator $\Ac[\cdot]$,
    and from Lemma~\ref{lemma:3} which shows that convergence holds
    also for the inverse operators $(\Ic-\Ac_n)^{-1}[\cdot]$. Then,
    using Lemma~\ref{lemma:4}, we show that
    $\phi_n(p)=(\Ic-\Ac_n)^{-1}[h_n](p)$ converges to
    $\phi(p)=(\Ic-\Ac)^{-1}[h](p)$.
  \end{itemize}
\endIEEEproof

\begin{lemma}\label{lemma:1}
Given the operator $\Ac[\cdot]$ defined above we have that
$\| \Ac[\phi](p)\|_{\mathrm{Lip}}<1$ for
opportunely chosen $c_0$ and $c_1$ under the assumption that
$\Phi(p,q)$ is $C^1(\Pc^2)$. Furthermore the operator
    $(\Ic-\Ac)^{-1}[\cdot]$ exists continuous with respect to norm $\| \cdot \|_{\rm Lip}$.
\IEEEproof
We  have that 
  \begin{eqnarray}
 |\Ac[\phi](p)| &\le &\int_q \left|\Phi(p,q) \phi(q)\rho_0(q) \dd q\right| 
  \le \int_q   |\Phi(p,q)\rho_0(q)|| \phi(q)| \dd q \non
  &\le &  \|\phi(p)\|_\infty \int_q   |\Phi(p,q)\rho_0(q)| \dd q
\end{eqnarray}
Therefore,  by assuming that $\alpha(p)$ and $\zeta(p,p')$ are regular in their
    argmmuments, we have 
\begin{eqnarray}\label{norm-inf}
  \| \Ac[\phi](p)\|_\infty &=& \sup_p \| \Ac[\phi](p)\| \non
  &\le& \|\phi(p)\|_\infty \sup_p  \int_q   \Phi(p,q)\rho_0(q) \dd q \non
  & \le &\|\phi(p)\|_\infty \frac{1}{\eta(p)}\sup_q \frac{\bar{\alpha}(q)\eta(q)}
            {\alpha(q) + \bar{\alpha}(q)\eta(q)} \underbrace{\int_q\zeta(p,q)\rho_0(q)\dd q}_{\eta(p)} \non
  &=& \|\phi(p)\|_\infty \sup_q \frac{\bar{\alpha}(q)\eta(q)}
      {\alpha(q) + \bar{\alpha}(q)\eta(q)} \non
  &=&  \kappa \|\phi(p)\|_\infty
\end{eqnarray}
where $\kappa<1$.
Similarly, using the Lagrange theorem, we have
\[  \sup_{p,q \in \Pc, p\neq q} \frac{|y(p)-y(q)|}{|p-q|} = \sup_p \left| \frac{\dd y(p) }{\dd p}\right| \]
for any continuous and differentiable function $y(p)$. Therefore, 
\begin{eqnarray}\label{norm-L}
\left| \frac{\dd \Ac[\phi](p) }{\dd p}\right| &= &\left| \int_q  \frac{\partial \Phi(p,q)}{\partial p} \phi(q) \rho_0(q)\dd q\right| \le\non
& \le &\int_q \left| \frac{\partial \Phi(p,q)}{\partial p} \phi(q)\rho_0(q) \right| \dd q\non
&\le & \left| \sup_{p,q\in \Pc} \frac{\partial \Phi(p,q)}{\partial p}\rho_0(q)\right| \|\phi(q)\|_\infty \nonumber\\
&=&   \| \Psi(p,q)\|_L  \|\phi(q)\|_\infty
\end{eqnarray}
Note that, to obtain the last expression, we defined $\|\Psi(p,q)\|_{\rm L} = \sup_{p,q\in \Pc}\left|\frac{\partial \Phi(p,q)}{\partial p} \rho_0(q)\right|$. 
 Now combining \eqref{norm-inf} and \eqref{norm-L}, we have:
 \[  \| \Ac[\phi](p)\|_{\rm Lip} \le c_0 \kappa { \|\phi(p)\|_\infty}+ c_1 \|\Psi(p,q)\|_{\rm L}  \|\phi(p)\|_\infty
\]
Dividing both sides by $\|\phi(p)\|_{\rm Lip}$ we get
\[
 \frac{ \| \Ac[\phi](p)\|_{\rm Lip}}{ \|\phi(p)\|_{\rm Lip}}\le c_0\frac{ \kappa { \|\phi(p)\|_\infty}}{ \|\phi(p)\|_{\rm Lip} }+ c_1\| \Psi(p,q)\|_{\rm L} \frac{  \|\phi(p)\|_\infty}{ \|\phi(p)\|_{\rm Lip}}.
\]
Now, since by construction $\frac{  \|\phi(p)\|_\infty}{\|\phi(p)\|_{\rm Lip}}\le \frac{1}{c_0}$, we have
\[ \|\Ac[\phi](p) \|_{\rm Lip} = \sup \frac{ \| \Ac[\phi](p)\|_{\rm Lip}}{ \|\phi(p)\|_{\rm Lip}}\le \kappa +\frac{c_1}{c_0}  \| \Psi(p,q)\|_{\rm L} 
\]
which can be made smaller than 1 by opportunely setting $c_0$ and $c_1$, i.e. by setting $\frac{c_1}{c_0} < \frac{1-\kappa}{\| \Psi(p,q)\|_{\rm L} }$.

For a generic linear operator $\Ac[\cdot]$ defined over a Banach space such that $\|\Ac[\cdot]\|<1$, the associated 
operator $(\Ic-\Ac)^{-1}[\cdot]$ exists continuous and can be written as~\cite[Th. 8 p. 102 and Th. 1 p. 111]{kolmogorov2012functions}
\begin{equation}\label{powerseries}
 (\Ic-\Ac)^{-1}[\cdot]= \sum_{k=0}^\infty (\Ac)^k[\cdot]\,.
\end{equation}
\endIEEEproof
\end{lemma}

We now consider the sequence of pertubated operators $\Ac_n[\cdot]$.
%such that % $\|\Ac_n[\cdot]\|_\infty<1$ and
%\[
% \|\Ac_n[\cdot] -\Ac[\cdot]\|_{\mathrm{Lip}} \to 0
%\]
The general result below applies. 
\begin{lemma}\label{lemma:2}
  Given a sequence of operators $\Ac_n[\cdot]$ and a sequence of
  functions $h_n(p)$ as defined above, they converge to, respectively,
  the above expressions of $\Ac[\cdot]$ and $h(p)$ in Lipschitz norm
  (i.e., $\|\Ac_n[\cdot]- \Ac[\cdot]\|_{\mathrm{Lip}} \to 0$ and
  $\|h_n(p)-h(p)\|_\mathrm{Lip}\to 0$).
\end{lemma}

\IEEEproof
To simplify the notation, without loss of generality we assume $\Pc=[0,1]$  (we  recall that $\Pc$ is assumed to be compact).
Then, 
\begin{eqnarray*}
\Ac_n[\phi(p)]= \int_\Pc   \Phi(p,q) \phi(q) \dd R_n(q)
 &= &\sum_{m=0}^{n-1 } \Phi\left(p,\frac{m}{n}\right) \phi\left(\frac{m}{n}\right) \int_{\frac{m}{n}}^{\frac{m+1}{n}} \dd R(q) 
\end{eqnarray*}
Furthermore, 
\begin{eqnarray*}
|\Ac[\phi(p)]-\Ac_n[\phi(p)]|
= \sum_{m=0}^{n-1} \int_{\frac{m}{n}}^{\frac{m+1}{n}} \left[\Phi(p,q)\phi(q)- \Phi\left(p,\frac{m}{n}\right) \phi\left(\frac{m}{n}\right)\right]\dd R(q)
\end{eqnarray*}
We therefore obtain: 
\begin{eqnarray*}
|\Ac[\phi(p)]-\Ac_n[\phi(p)]|&\le & \sum_{m=0}^{n-1}\int_{\frac{m}{n}}^{\frac{m+1}{n}} \left| \Phi(p,q)\right|\left|\phi(q)- \phi\left(\frac{m}{n}\right)\right| \dd R(q)\\
&+& \sum_{m=0}^{n-1}\int_{\frac{m}{n}}^{\frac{m+1}{n}} \left| \Phi(p,q)-\Phi\left(p,\frac{m}{n}\right)\right| \left| \phi\left(\frac{m}{n}\right) \right| \dd R(q)\\
&\le& \|\Phi(p,q)\|_\infty\frac{ \| \phi(p)\|_L}{n}\int_0^1 \dd R(q)\\
&+& \frac{\|\Phi(p,q)\|_{\rm L}}{n} \|\phi(p)\|_\infty\int_0^1 \dd R(q)\\
                             &=& \frac{1}{n}\left(\|\Phi(p,q)\|_\infty \|\phi(p)\|_L+\|\Phi(p,q)\|_L \|\phi(p)\|_\infty\right)
\end{eqnarray*}

It follows that:
\[
\|\Ac[\phi(p)]-\Ac_n[\phi(p)]\|_\infty\le \frac{1}{n}\left(\|\Phi(p,q)\|_\infty \| \phi(p)\|_L+\|\Phi(p,q)\|_L \|\phi(p)\|_\infty\right)
\]

Similarly, since $\Phi(p,p')$ is assumed to be differentiable with
continuous derivative with respect to $p$ and $q$, both $\Ac[\phi(p)]$
and $\Ac_n[\phi(p)]$ are differentiable at every point with continuous
derivative and:
\begin{eqnarray*}
\left|  \frac{\dd \Ac[\phi(p)]}{\dd p}-  \frac{\dd \Ac_n[\phi(p)]}{\dd p}\right|&=&
\left|\sum_{m=0}^{n-1} \int_{\frac{m}{n}}^{\frac{m+1}{n}} \left[\frac{\partial \Phi(p,q)}{\partial p}\phi(q)- 
\frac{\partial \Phi\left(p,\frac{m}{n}\right)}{\partial p} \phi\left(\frac{m}{n}\right)\right] \dd R(q)\right|
\end{eqnarray*}
Proceeding as before, we get:
\begin{eqnarray*} 
  \|\Ac[\phi(p)]-\Ac_n[\phi(p)] \|_L  &\le& \frac{1}{n}\left( \left\| \frac{\partial \Phi(p,q)}{\partial p}  \right\|_\infty  \| \phi(p) \|_L +\left\| \frac{\partial \Phi(p,q)}{\partial p} \right\|_L  \| \phi(p)\|_\infty  \right)
\end{eqnarray*}
and
\begin{eqnarray*} 
  \|\Ac[\phi(p)]-\Ac_n[\phi(p)]\|_{\mathrm{Lip}}
  &\le& 
   \frac{c_0}{n}\left( \|\Phi(p,q)\|_\infty\| \phi(p)\|_L
+ \Phi(p,q)\|_L \|\phi(p)\|_\infty  \right) \\
&&  + \frac{c_1}{n}\left(\left\| \frac{\partial \Phi(p,q)}{\partial p} \right\|_\infty  \| \phi(p)\|_L
  + \left\|\frac{\partial \Phi(p,q)}{\partial p}\right\|_L \|\phi(p)\|_\infty \right)
\end{eqnarray*}
Since in the right hand side of the above expressions none of the norms depend on $n$, it easy to see that  
$\| \Ac[\phi(p)] - \Ac_n[\phi(p)]\|_{\mathrm{Lip}}\to 0$ as $n\to \infty$. 
With similar arguments, we can prove  that $\|h(p)-h_n(p)\|_{\mathrm{Lip}}\to 0$ as $n\to \infty$.
\endIEEEproof

\begin{lemma}\label{lemma:3}
Given a Banach space with norm $\|\cdot \|$, and
 a sequence of linear operators $\Ac_n[\cdot] \to \Ac[\cdot]$ in norm,  with $\|\Ac[\cdot]\|<1$, 
we  have that the continuous operators
$(\Ic[\cdot]-\Ac_n[\cdot] )^{-1}\to (\Ic[\cdot]-\Ac [\cdot] )^{-1}$ in norm.
\end{lemma}
\IEEEproof Given that $\|\Ac[\cdot]\|<1$ by the continuity of norm
$\|\Ac_n[\cdot]\|\to \|\Ac[\cdot]\|<1$, for $n$ sufficiently large we
can assume $\|\Ac_n[\cdot]\|<1$.  For any of such $n$, we define
$\Bc_n[\cdot]=(\Ic[\cdot]-\Ac_n[\cdot])^{-1}- (\Ic[\cdot]-\Ac[\cdot])^{-1}$. By \eqref{powerseries}, we can write
\[
 \Bc_n[\cdot]=\sum_{k=0}^{\infty} (\Ac_n[\cdot])^k- \sum_k  (\Ac[\cdot])^k\,.
\]
Since both series on the right hand side  of the above expression
converge, we can write:
\[
 \Bc_n[\cdot]=\sum_{k=0}^{\infty} (\Ac_n[\cdot])^k-  (\Ac[\cdot])^k
\]

Now,  we have:

\[
\| \Bc_n[\cdot]\| = \left\|\sum_{k=0}^{\infty} (\Ac_n[\cdot])^k-  (\Ac[\cdot])^k\right\|\le \sum_{k=0}^{\infty}  \|(\Ac_n[\cdot])^k-  (\Ac[\cdot])^k\|
\]
where the last inequality follows by the sub-additivity and continuity of the norm.

Denoted with
$c_k=\|(\Ac_n[\cdot])^k- (\Ac[\cdot])^k\|=
\|[(\Ac_n[\cdot]-\Ac[\cdot])+\Ac[\cdot]]^k- (\Ac[\cdot])^k\|$,
and considering that the operator algebra is, in general, non-commutative, we
have:
\begin{eqnarray*}
  c_k
  &= &\left\| \sum_{\substack{x_i \in \{0,1\}, i=1,\ldots,k}} \prod_i (\Ac[\cdot])^{x_i}(\Ac_n[\cdot]-\Ac[\cdot])^{1-x_i}-(\Ac[\cdot])^k \right\| \\ 
 &= &\left\| \sum_{\substack{x_i \in \{0,1\}, i=1,\ldots,k\\ \{x_1,x_2\cdots x_k \}\neq \{1,1,\cdots,1\} }}\prod_i (\Ac[\cdot])^{x_i}(\Ac_n[\cdot]-\Ac[\cdot])^{1-x_i} \right\|\\ 
 &\le &   \sum_{\substack{x_i \in \{0,1\}, i=1,\ldots,k\\ \{x_1,x_2\cdots x_k \}\neq \{1,1,\cdots,1\} }} \prod_i \| (\Ac[\cdot])^{x_i}\| \|(\Ac_n[\cdot]-\Ac[\cdot])^{1-x_i}\| \nonumber\\ 
    & = & \sum_{i=1}^k {k \choose i}\|\Ac_n[\cdot]-\Ac[\cdot]\|^{i}\|\Ac[\cdot]\|^{k-i}\nonumber\\ 
 &=& \sum_{i=1}^k {k \choose i}\|\Ac[\cdot]\|^i\|\Ac_n[\cdot]-\Ac[\cdot]\|^{k-i}\nonumber\\ 
     &=& (\|\Ac[\cdot]\| + \|\Ac_n[\cdot]-\Ac[\cdot]\|)^k- \|\Ac[\cdot]\|^k\,.
\end{eqnarray*}
Therefore, by monotonicity of positive series, we have:
\begin{equation}
  \| \Bc_n[\cdot]\| \le \sum_{k=0}^\infty c_k %\le \sum_{k=0}^\infty c_k
                     \leq  \sum_{k=0}^\infty (\|\Ac[\cdot]\|+\|\Ac_n[\cdot]-\Ac[\cdot]\|)^k- \|\Ac[\cdot]\|^k
\end{equation}
%Now by construction $\sum_k |\|\Ac[\cdot]\|_\infty^k- \|\Ac_n[\cdot]\|^k|= \left|\sum_k [\|\Ac[\cdot]\|_\infty^k- \|\Ac_n[\cdot]\|^k_\infty]\right|$, 
%since  the term of the last series are all of the same sign by construction. 
Since $ \|\Ac_n[\cdot]-\Ac[\cdot]\|_\infty\to 0$ and   $\|\Ac[\cdot]\|_\infty<1$, we can assume $n$ sufficiently large so that  $\|\Ac[\cdot]\|+
\|\Ac_n[\cdot]-\Ac[\cdot]\|<1$, thus:
\begin{eqnarray*}
 \| \Bc_n[\cdot]\| &\le&  \sum_{k=0}^\infty (\|\Ac[\cdot]\|+\|\Ac_n[\cdot]-\Ac[\cdot]\|)^k- \|\Ac[\cdot]\|^k \\
& = &  \left| \sum_{k=0}^\infty (\|\Ac[\cdot]\|+\|\Ac_n[\cdot]-\Ac[\cdot]\|)^k- \sum_{k=0}^\infty \|\Ac_n[\cdot]\|^k\right|\\
&=&  \left|\frac{1}{1- (\|\Ac[\cdot]\|+\|\Ac_n[\cdot]-\Ac[\cdot]\|)}- \frac{1}{1- \|\Ac[\cdot]\|}\right|
\end{eqnarray*}
The thesis follows immediately since $ \|\Ac_n[\cdot]-\Ac[\cdot]\|\to 0$, hence $ \| \Bc_n[\cdot]\|\to 0$.

\endIEEEproof

\begin{lemma}\label{lemma:4}
Given  the Banach space  $\mathrm{Lip}(\Pc)$, 
a sequence of linear and continuous operators $\Cc_n[\cdot]$: $\mathrm{Lip}[\Pc]\to \mathrm{Lip}[\Pc]$  
converging in $\| \cdot \|_{\mathrm{Lip}}$-norm to  the continuous operator $\Cc$,
and  a sequence of  functions $y_n \in \mathrm{Lip}(\Pc)$ converging to $y$  in $\| \cdot \|_{\mathrm{Lip}}$-norm, then
$\Cc_n[y_n] $  converges in $\| \cdot \|_{\mathrm{Lip}}$-norm to $\Cc[y]$. 
\end{lemma}
\IEEEproof

To simplify the notation, we denote $\| \cdot \|_{\mathrm{Lip}}$ simply with  $\|\cdot\|$: 
\begin{eqnarray*}
\|\Cc_n[y_n] - \Cc[y]\| & = & \|\Cc_n[y_n] -\Cc_n[y_0] +  \Cc_n[y_0]- \Cc[y]\|\\
& \le  &  \|\Cc_n[y_n- y_0]\|+ \| \Cc_n[y]- \Cc[y] \|
\end{eqnarray*}
 Now,  on the one hand, $\|\Cc_n[y_n- y] \| \le \|\Cc_n[\|y_n-y\|] \| \to 0$ given that $\|\Cc_n[\cdot]\|$ is bounded since it is continuous, and $\|y_n -y\|\to 0$.
 On the other hand, $\| \Cc_n[y]- \Cc[y] \|  \to 0$ since  $\| \Cc_n[\cdot]- \Cc[\cdot] \| \to 0$,  while $y$ is bounded in norm. The assertion follows immediately.

\endIEEEproof
\end{document}